\documentclass[prl,twocolumn,aps,superscriptaddress]{revtex4}
\usepackage{amsfonts}

\usepackage{dcolumn}
\usepackage{bm}
\usepackage{tikz}
\usepackage[colorlinks=true,citecolor=blue,urlcolor=blue]{hyperref}
\usepackage{amsmath,amsfonts,amssymb,times,natbib}
\usepackage[standard]{ntheorem}
\usepackage{graphicx}
\usepackage{subfigure}

\begin{document}

%\title{Experimental Test of the Einstein-Podolsky-Rosen Paradox and its Generated Steering Inequality}
%\title{\textcolor{blue}{A Steering Paradox ``k=1'' and its Generated Inequality}}
%\title{A Generalized Quantum Paradox for Einstein-Podolsky-Rosen Steering and its Extended Inequality}
\title{A Steering Paradox for Einstein-Podolsky-Rosen Argument and its Extended Inequality}

\author{Tianfeng Feng}
\affiliation{State Key Laboratory of Optoelectronic Materials and Technologies and School of Physics, Sun Yat-sen University, Guangzhou, People's Republic of China}

\author{Changliang Ren}
\email{renchangliang@cigit.ac.cn}
\affiliation{Key Laboratory of Low-Dimensional Quantum Structures and Quantum Control of Ministry of Education, Key Laboratory for Matter
Microstructure and Function of Hunan Province, Department of Physics and Synergetic Innovation Center for Quantum Effects and
Applications, Hunan Normal University, Changsha 410081, China}

\author{Qin Feng}
\affiliation{State Key Laboratory of Optoelectronic Materials and Technologies and School of Physics, Sun Yat-sen University, Guangzhou, People's Republic of China}

\author{Maolin Luo}
\affiliation{State Key Laboratory of Optoelectronic Materials and Technologies and School of Physics, Sun Yat-sen University, Guangzhou, People's Republic of China}

\author{Xiaogang Qiang}
\affiliation{National Innovation Institute of Defense Technology, AMS, Beijing, China}

\author{Jing-Ling Chen}
\email{chenjl@nankai.edu.cn}
\affiliation{Theoretical Physics Division, Chern Institute of Mathematics, Nankai University, Tianjin 300071, People's Republic of China}

\author{Xiaoqi Zhou}
\email{zhouxq8@mail.sysu.edu.cn}
\affiliation{State Key Laboratory of Optoelectronic Materials and Technologies and School of Physics, Sun Yat-sen University, Guangzhou, People's Republic of China}

\date{\today}
\begin{abstract}
The Einstein-Podolsky-Rosen (EPR) paradox is one of the milestones in quantum foundations, arising from the lack of local realistic description of quantum mechanics. The EPR paradox has stimulated an important concept of ``quantum nonlocality", which manifests itself by three different types: quantum entanglement, quantum steering, and Bell nonlocality. Although Bell nonlocality is more often used to show the ``quantum nonlocality'', the original EPR paradox is essentially a steering paradox. In this work, we formulate the original EPR steering paradox into a contradiction equality, %``$k=1$"
thus making it amenable to an experimental verification. We perform an experimental test of the steering paradox in a two-qubit scenario. Furthermore, by starting from the steering paradox, we generate a generalized linear steering inequality and transform this inequality into a mathematically equivalent form, which is more friendly for experimental implementation, i.e., one may only measure the observables in $x$-, $y$-, or $z$-axis of the Bloch sphere, rather than other arbitrary directions. We also perform experiments to demonstrate this scheme. Within the experimental errors, the experimental results coincide with the theoretical predictions. Our results deepen the understanding of quantum foundations and provide an efficient way to detect the steerability of quantum states.
%\textcolor{blue}{Einstein-Podolsky-Rosen steering, as a quantum resource, is a form of novel nonlocality intermediate between quantum entanglement and Bell nonlocality. So far, there are many steering criterions based on certain statistical correlations determined by measurements on a two-particle systems, but these criterions  based on logical contradictions or paradoxes are poor. Here we report an experimental verification of a generalized quantum steering paradox ``$k=1$'' which is stronger than the previous ones in showing the conflict of quantum mechanics with local hidden state model. Furthermore, from the paradox we can generate the generalized linear steering inequalities, which has an advantage over the usual one by detecting more steerability of quantum states. Besides, the inequalities can be transformed into a mathematically equivalent form, but is more friendly for experimental implementation, i.e., one may only measure the observables in $x$-, $y$-, or $z$-axis, rather than other arbitrary directions. We also perform experiments to confirm this merit based on photonic system and show that within the experimental errors, the experimental results coincide with the theoretical predictions. Our results provide a novel way to detect quantum steering and deepen the understanding of quantum correlation and quantum foundations.}% we experimentally observed the stronger quantum paradox and violations of the inequality in photonic system.
 %At last, we experimentally observed the stronger quantum paradox and violations of the inequality in photonic system.
\end{abstract}


\maketitle

\section{1. Introduction}
Quantum paradox has provided an intuitive way to reveal the essential difference between quantum mechanics and classical theory. In 1935, by considering a continuous-variable entangled state $\Psi(x_1, x_2)=\int_{-\infty}^{+\infty} e^{i p(x_1-x_2+x_0)/\hbar}dp$, Einstein, Podolsky and Rosen (EPR) proposed a thought experiment to highlight a famous paradox~\cite{EPR}: either the quantum wave-function does not provide a complete description of physical reality, or measuring one particle from a quantum entangled pair instantaneously affects the second particle regardless of how far apart the two entangled particles are. The EPR paradox has revealed a sharp conflict between local realism and quantum mechanics, thus triggering the investigation of nonlocal properties of quantum entangled states. Soon after the publication of the EPR paper, Schr{\" o}dinger made an immediate response by introducing the term ``steering" to depict ``the spooky action at a distance" that was mentioned in the EPR argument~\cite{Schrodinger35}. According to Schr{\" o}dinger, ``steering" reflects a nonlocal phenomenon which, in a bipartite scenario, describes the ability of one party, say Alice, to prepare the other party (say Bob) particle in different quantum states by simply measuring her own particle using different settings. However, the notion of steering has not been gained much attention and development until 2007, when Wiseman \emph{et al.} gave a rigorous definition using concepts from quantum information~\cite{wiseman07}.

Undoubtedly, the EPR paradox is a milestone in quantum foundations, for it has opened the door of ``quantum nonlocality". In 1964, Bell made a distinct response to the EPR paradox by showing that quantum entangled states may violate Bell inequality, which hold for any local-hidden-variable model \cite{Bell}. This indicates that the local-hidden-variable models cannot reproduce all quantum predictions, and the violation of Bell inequality by entangled states directly implies that a kind of nonlocal properties -- Bell nonlocality. Since then, Bell nonlocality has been achieved a rapid and fruitful development in two directions~\cite{BrunnerRMP}: (i) On one hand, more and more Bell's inequalities have been introduced to detect Bell's nonlocality in different physical systems, for examples, the Clause-Horne-Shimony-Holt (CHSH) inequality for two qubits~\cite{CHSH}, the Mermin-Ardehali-Belinskii-Klyshko (MABK) inequality for multipartite qubits~\cite{MABK}, the Collins-Gisin-Linden-Masser-Popescu inequality for two qudits~\cite{CGLMP}. (ii) On the other hand, some novel quantum paradoxes, or the all-versus-nothing (AVN) proofs have been suggested to reveal Bell's nonlocality without inequalities. Typical examples are the Greenberger-Horne-Zeilinger (GHZ) paradox~\cite{GHZ} and the Hardy paradox~\cite{Hardy}. Experimental verifications of Bell nonlocality have also been carried out, for instances, Aspect \emph{et al.} have successfully made the first observation of Bell's nonlocality with the CHSH inequality~\cite{Aspect}, Pan \emph{et al.} have tested the three-qubit GHZ paradox in the photon-based experiment~\cite{Pan}, and very recently Luo \emph{et al.} have tested the generalized Hardy paradox for the multi-qubit systems~\cite{Luo}.

Despite of being developed from the EPR paradox, Bell nonlocality does not directly correspond to the EPR paradox.
As pointed out in Ref.~\cite{wiseman07}, inspired by the EPR argument, one can derive three different types of ``quantum nonlocality" : quantum entanglement, quantum steering, and Bell nonlocality. %
%and quantum steering is subset of quantum entanglement while it is superset of Bell nonlocality. %exactly the type of ``quantum nonlocality" inherited in the EPR paradox.
The original EPR paradox is actually a special case of quantum steering \cite{Reid}.
Although quantum steering has been experimentally demonstrated in various quantum systems~\cite {saunders10,sun14,armstrong15,Bennet,Schneeloch,Fadel,Dolde,Dabrowski,Cavalcanti, Zeng}, all of these experiments just indirectly illustrate the EPR paradox, in which most of them are based on statistical inequalities. Here the direct illustration of a quantum paradox means that we can find a contradiction equality for this paradox and demonstrate it (the ref. \cite{sun14} is a AVN proof but not a contradiction equality). 
%Since 2007, theoretically and experimentally quantum steering has been demonstrated by the violation of EPR steering inequalities~\cite{saunders10} as well as the AVN proof~\cite{chenavn}.
%has not yet been tested experimentally since 1935, although it is the earliest quantum paradox appeared in the literature.
%There are probably two main reasons: (i) The original EPR state $\Psi(x_1, x_2)$ is an entangled state of bipartite continuous-variable system, which is challenging to prepare in the lab. (ii) The ``paradox" in the EPR argument is not very clear in its mathematical formulation, and correspondingly, it is difficult to understand what should be observed or measured experimentally.
%To directly illustrate a quantum paradox, a well-formulated contradiction equality is required. 
For examples, (i) The GHZ paradox \cite{GHZ} can be formulated as a contradiction equality ``$+1=-1$", where ``$+1$" represents the prediction of the local-hidden-variable model, while ``$-1$" is the quantum prediction. Thus, if one observes the value of ``$-1$"  by some quantum technologies in the experiments, then the GHZ paradox is demonstrated. (ii) The formulation of the Hardy paradox \cite{Hardy} is given as follows: under some certain Hardy-type constraints for probabilities $P_1=P_2=\cdots=P_N=0$, any local-hidden-variable model predicts a zero-probability (i.e., $P_{\rm suc}=0$), while quantum prediction is $P_{\rm suc}>0$, where $P_{\rm suc}$ is the success probability of a specific event. Upon successfully measuring the desired non-zero success probability under the required Hardy constraints, one verifies the Hardy paradox. 
 A natural question arises whether the EPR paradox, which excludes any local-hidden-state (LHS) model, can be illustrated in a direct way just like GHZ or Hardy paradox? 

%A natural question arises whether the EPR paradox can be illustrated in a direct way just like GHZ or Hardy paradox?
%In fact, one of our authors has tried to answer this question before~\cite{chen16}, but the result is only a special case and no experimental verification.}

%In the literature, as two well-known quantum paradoxes related to Bell's nonlocality, the GHZ paradox and the Hardy paradox have been widely studied and also experimentally confirmed. In addition, a quantum paradox can correspond to an inequality. For examples, (i) The two-qubit Hardy paradox may correspond to the well-known CHSH inequality by the following equivalent construction: $P_{\rm suc}-(P_1+P_2+P_3)\leq 0$, and (ii) The three-qubit GHZ paradox may correspond to the three-qubit MABK inequality. In Ref.~\cite{chen16}, the researchers have shown that, for any bipartite entangled pure state (including the original EPR state), the original EPR paradox can be mathematically formulated into a contradiction equality ``$k=1$", thus making it amenable to an experimental verification.

The purpose of this paper is two-fold: (i) Based on our previous results of the steering paradox ``$2=1$'' \cite{chen16}, we present a generalized steering paradox ``$k=1$''. We have also performed an experiment to illustrate the original EPR paradox through demonstrating the steering paradox ``$2=1$'' in a two-qubit scenario. (ii) A steering paradox can correspond to an inequality (e.g., the two-qubit Hardy paradox may correspond to the well-known CHSH inequality)~\cite{Mermin94,chen18}, and from the steering paradox ``$k=1$'', we generate a generalized linear steering inequality (GLSI), which naturally includes the usual linear steering inequality as a special case \cite{wiseman07,saunders10}. Besides, the GLSI can be transformed into a mathematically equivalent form, but is more friendly for experimental implementation, i.e., one may only measure the observables in $x$-, $y$-, or $z$-axis of the Bloch sphere, rather than other arbitrary directions. Meanwhile, we also experimentally test quantum violations of the GLSI, which shows that it is more powerful than the usual one in detecting the steerability of quantum states.

% \begin{figure}
%\includegraphics[width=0.4\textwidth]{1.eps}
%\caption{Scenario of quantum steering. Alice shares a bipartite quantum states and send one particle to Bob. And then, they measure their respective particle and communicate classically. Bob can verify the whole process whether it indicates a steering phenomenon or not via calculating correlation value $S$ between Alice's and his particle. }
%\end{figure}

\section{2. The EPR paradox as a steering paradox ``$k=1$"}
%In 2007, Wiseman \emph{et al.} have classified quantum nonlocality into three different types: quantum entanglement, EPR steering, and Bell's %nonlocality~\cite{wiseman07}.
%By presenting the steering paradox ``$k=1$", Ref.~\cite{chen16}  has successfully confirmed that EPR steering is exactly the type of ``quantum %nonlocality" inherited in the EPR paradox.
Following~Ref.\cite{chen16}, let us consider an arbitrary two-qubit pure entangled state $\rho_{AB}=|\Psi(\alpha,\varphi)\rangle\langle\Psi(\alpha,\varphi)|$ shared by Alice and Bob. Using the Schmidt decomposition, i.e., in the $\hat{z}$-direction representation, the wave-function $|\Psi(\alpha,\varphi)\rangle$ may be written as
\begin{eqnarray}\label{decom-z}
 |\Psi(\alpha,\varphi)\rangle=\cos\alpha|00\rangle+e^{i\varphi}\sin\alpha|11\rangle,
 \end{eqnarray}
with $\alpha\in (0, \pi/2), \varphi \in[0, 2\pi]$. For the same state Eq. (\ref{decom-z}), in the general $\hat{n}$-direction decomposition one may recast it to
\begin{eqnarray}\label{decom-n}
 |\Psi(\alpha,\varphi)\rangle &=& |+\hat{n}\rangle |\chi_{+\hat{n}}\rangle+ |-\hat{n}\rangle |\chi_{-\hat{n}}\rangle,
 \end{eqnarray}
where $|\pm \hat{n}\rangle$ are the eigenstates of the operator $\hat{P}_a^{\hat{n}}=[\openone+(-1)^a \vec{\sigma}\cdot \hat{n}]/2$
denoting Alice's projective measurement on her qubit along the $\hat{n}$-direction with measurement outcomes $a$ ($a=0, 1$), $\openone$ is the identity matrix , $\vec{\sigma}=(\sigma_x, \sigma_y, \sigma_z)$ is the vector of Pauli matrices, and $|\chi_{\pm\hat{n}}\rangle=\langle \pm \hat{n}|\Psi(\alpha,\varphi)\rangle$ are the collapsed pure states (unnormalized) for Bob's qubit.

By performing a projective measurement on her qubit along the $\hat{n}$-direction, Alice, by wavefunction collapse, steers Bob's qubit to the pure states ${\rho}^{\hat{n}}_a=\tilde{\rho}^{\hat{n}}_a/{\rm
tr}(\tilde{\rho}^{\hat{n}}_a)$ with the probability ${\rm
tr}(\tilde{\rho}^{\hat{n}}_a)$, here $\tilde{\rho}^{\hat{n}}_a={\rm tr}_A[(\hat{P}_a^{\hat{n}}\otimes \openone)\;\rho_{AB}]$ are the so-called Bob's unnormalized conditional states and ${\rho}^{\hat{n}}_a$ are the normalized ones~\cite{wiseman07}.
%The pure entangled state $|\Psi\rangle$ has a remarkable property: Bob's normalized conditional states
%are always \emph{pure}, and ${\rho}^{\hat{n}_1}_a\neq {\rho}^{\hat{n}_2}_a$ if $\hat{n}_1\neq \hat{n}_2$ .
In a two-setting steering protocol $\{\hat{z}, \hat{x}\}$, if Bob's four unnormalized conditional states can be simulated by an ensemble $\{\wp_\xi \rho_\xi\}$ of the LHS model, then these may be described as \cite{chen16}
% (see the Methods in~\cite{chen16})
\begin{subequations}  \label{E0-p4}
\begin{eqnarray}
 \tilde{\rho}^{\hat{z}}_0&=& \cos^2\alpha|0\rangle\langle 0|= \wp_1 \rho_1,\label{Erhoz0q-4}\\
 \tilde{\rho}^{\hat{z}}_1&=& \sin^2\alpha|1\rangle\langle 1|=\wp_2 \rho_2,\label{Erhoz1q-4}\\
 \tilde{\rho}^{\hat{x}}_0&=& (1/2)|\chi_+\rangle\langle\chi_+|=\wp_3 \rho_3,\label{Erhox0q-4}\\
 \tilde{\rho}^{\hat{x}}_1&=& (1/2)|\chi_-\rangle\langle\chi_-|=\wp_4 \rho_4,\label{Erhox1q-4}
\end{eqnarray}
\end{subequations}
where $|\chi_\pm\rangle =\cos\alpha|0\rangle\pm e^{i\varphi}\sin\alpha|1\rangle$ are normalized pure states, the $\rho_i$ are hidden states, and the $\wp_i$ represent the corresponding probabilities in the ensemble. They satisfy the constraint  $\sum_\xi \wp_\xi \rho_\xi=\rho_B={\rm tr}_A [\rho_{AB}]$, where $\rho_B$ is the reduced density matrix of Bob. On the other hand, since $\tilde{\rho}^{\hat{n}}_0+\tilde{\rho}^{\hat{n}}_1=\rho_B$ and ${\rm tr}\rho_B=1$, if we sum up terms in Eq. (\ref{E0-p4}) and take the trace, we arrive at the contradiction ``$2=1$'', which represents the EPR paradox in the two-setting steering protocol.

Here we show that a more general steering paradox ``$k=1$'' can be similarly obtained  if one considers a $k$-setting steering scenario $\{\hat{n}_1, \hat{n}_2, \cdots, \hat{n}_k \}$, in which Alice performs $k$ projective measurements on her qubit along $\hat{n}_j$-directions (with $j=1, 2,\cdots, k$).  For each projective measurement $ \hat {P}^{\hat{n}_j}_a$,
Bob obtains the corresponding unnormalized pure states $\tilde{\rho}^{\hat{n}_j}_a$. Suppose these states can be simulated by the LHS model, then one may obtain the following set of $2k$ equations:
\begin{subequations}  \label{E0-pk}
\begin{eqnarray}
 \tilde{\rho}^{\hat{n}_j}_0&=&\sum_\xi \wp(0|\hat{n}_j,\xi) \wp_{\xi} \rho_{\xi},\label{Erhonj0q-2}\\
 \tilde{\rho}^{\hat{n}_j}_1&=&\sum_\xi \wp(1|\hat{n}_j,\xi) \wp_{\xi} \rho_{\xi}, \;\;\;(j=1, 2,\cdots, k).\label{Erhoj1q-2}
% && j=1,2,\cdots, k. \nonumber
\end{eqnarray}
\end{subequations}
Since the $\tilde{\rho}^{\hat{n}_j}_a$'s are proportional to pure states, the sum of the right-hand side of Eq. (\ref{E0-pk}) actually contains only one $\rho_{\xi}$, as we have seen for Eq. (\ref{E0-p4}). Furthermore, due to the relations
$\tilde{\rho}^{\hat{n}_j}_0+  \tilde{\rho}^{\hat{n}_j}_1=\rho_B$ and $\sum_{\xi=1}^{2k} \wp_{\xi} \rho_{\xi}=\rho_B$,
%\begin{eqnarray}
%\tilde{\rho}^{\hat{n}_j}_0+  \tilde{\rho}^{\hat{n}_j}_1=\rho_B, \;\;\; \sum_{\xi=1}^{2k} \wp_{\xi} \rho_{\xi}=\rho_B,
%\end{eqnarray}
and by taking trace of  Eq. (\ref{E0-pk}), one immediately has the steering paradox ``$k=1$''.

Experimentally, we shall test the EPR paradox for a two-qubit system in the simplest case of $k=2$. To this aim, we need to perform measurements leading to four quantum probabilities. The first one is  $P^{\rm QM}_1={\rm tr} [ \tilde{\rho}^{\hat{z}}_0\;|0\rangle\langle 0|]=\cos^2\alpha$, which is obtained from Bob by performing the projective measurement $|0\rangle\langle 0|$ on his unnormalized conditional state as in Eq. (\ref{Erhoz0q-4}). Similarly, from Eq. (\ref{Erhoz1q-4})-Eq. (\ref{Erhox1q-4}), one has $P^{\rm QM}_2={\rm tr} [ \tilde{\rho}^{\hat{z}}_1\;|1\rangle\langle 1|]=\sin^2\alpha$, $P^{\rm QM}_3={\rm tr} [ \tilde{\rho}^{\hat{x}}_0\;|\chi_+\rangle\langle\chi_+|]=1/2$, and $P^{\rm QM}_4={\rm tr} [ \tilde{\rho}^{\hat{x}}_1\;|\chi_-\rangle\langle\chi_-|]=1/2$. Consequently, the total quantum prediction is $P^{\rm QM}_{\rm total}=\sum_{i=1}^{4} P^{\rm QM}_i=2$,
which contradicts the LHS-model prediction ``1''. If, within the experimental measurement errors, one obtains a value $P^{\rm QM}_{\rm total}\approx 2$, then the steering paradox ``$2=1$'' is demonstrated.

\section{Generalized linear steering inequality} Just like Bell's inequalities may be derived from the GHZ and Hardy paradoxes~\cite{Mermin94,chen18}, this is also the case for the EPR paradox. In turn, from the steering paradox ``$k=1$'', one may derive a $k$-setting generalized linear steering inequality as follows: In the steering scenario $\{\hat{n}_1, \hat{n}_2, \cdots, \hat{n}_k \}$, Alice performs $k$ projective measurements along $\hat{n}_j$-directions. Upon preparing the two-qubit system in the pure state $|\Psi(\theta,\phi)\rangle$ (note that this is not $|\Psi(\alpha,\varphi)\rangle$ and here $|\Psi(\theta,\phi)\rangle$ is used to derive the inequality), for each measurement $\hat{P}_a^{\hat{n}_j}$, Bob has the corresponding normalized pure states as ${\rho}^{\hat{n}_j}_a(\theta,\phi)=\tilde{\rho}^{\hat{n}_j}_a/{\rm
tr}(\tilde{\rho}^{\hat{n}_j}_a)$, where $\tilde{\rho}^{\hat{n}_j}_a={\rm tr}_A[(\hat{P}_a^{\hat{n}_j}\otimes \openone)\;|\Psi(\theta,\phi)\rangle\langle \Psi(\theta,\phi)|]$ with ($a=0, 1$). Then the $k$-setting GLSI is given by (see Appendix A)
\begin{eqnarray}\label{k-setting inequality m}
\mathcal{S}_k(\theta,\phi)&=& \sum_{j=1}^k \biggr(\sum_{a=0}^1 P(A_{n_j}=a)\;\langle  {\rho}^{\hat{n}_j}_a(\theta,\phi) \rangle \biggr)\\&&
\leq C_{\rm LHS},
\end{eqnarray}
which is a $(\theta,\phi)$-dependent inequality, where $C_{\rm LHS}$ is the classical bound determined by the maximal eigenvalue of  $\mathcal{S}_k(\theta,\phi)$ for the given values of $\theta$ and $\phi$,   $P(A_j=a)$ is the probability of the $j$-th measurement of Alice with outcome $a$, and ${\rho}^{\hat{n}_j}_a(\theta,\phi)=|\chi^j_\pm(\theta,\phi)\rangle\langle \chi^j_\pm(\theta,\phi)|$
corresponds to Bob's projective measurements. This inequality can be used to detect the steerability of two-qubit pure or mixed states.

%The generated steering inequality (\ref{k-setting inequality}) is a $(\theta,\phi)$-dependent inequality, which may be used to detect the steerability of two-qubit quantum states (pure or mixed). For the pure states  given in Eq. (\ref{decom-z}), one has maximal violation, i.e. $\langle\Psi|\mathcal{S}_k|\Psi\rangle=k$ by choosing $\theta=\alpha$, $\phi=\varphi$, as it follow from the relation $\sum_{a=0}^1{\rm tr}[(\hat{P}_a^{\hat{n}_j}\otimes {\rho}^{\hat{n}_j}_a) |\Psi\rangle\langle\Psi|]=\sum_{a=0}^1{\rm tr}(\tilde{\rho}^{\hat{n}_j}_a)=1$.

The GLSI has two remarkable advantages over the usual LSI~\cite{saunders10}: (i) Based on its own form as in Eq. (\ref{k-setting inequality m}), the GLSI includes naturally the usual LSI as a special case, thus can detect more quantum states. In particular, the GLSI can detect the steerability for all pure entangled states Eq. (\ref{decom-z}) in the whole region $\alpha\in (0, \pi/2)$, at variance with the usual LSI, which fails to detect EPR steering for some regions of $\alpha$ close to 0 \cite{X}. (ii) The use of GLSI reduces the numbers of experimental measurements and improves the experimental accuracy. This may be seen as follow: with the usual $k$-setting LSI, Bob needs to perform $k$ measurements in different $k$ directions, for different input states $\rho_{AB}$. This is experimentally challenging since it may be hard to suitably tune the setup for all the $k$ directions. However, with the GLSI one may solve this issue using the Bloch realization $|\chi^j_\pm\rangle\langle \chi^j_\pm|=(\openone+\vec{\sigma}\cdot\hat{m}^j_\pm)/2$, which transforms the GLSI to an equivalent form where Bob only needs to perform measurements along the $\hat{x}$, $\hat{y}$ and $\hat{z}$ directions of the Bloch sphere, which are independent on the input states~(see Appendix A).

To be more specific, we give an example of the 3-setting  GLSI from Eq. (\ref{k-setting inequality m}), where Alice's three measuring directions are $\{\hat{x}, \hat{y}, \hat{z}\}$. Then we immediately have
\begin{eqnarray}\label{3-setting inequality}
\mathcal{S}_3&=&P(A_x=0)\; \langle
 |\chi_+\rangle\langle \chi_+|
\rangle+P(A_x=1)\;\langle |\chi_-\rangle\langle \chi_-| \rangle
\nonumber\\
&&+ P(A_y=0)\; \langle
 |\chi'_+\rangle\langle \chi'_+|
\rangle+P(A_y=1)\;\langle |\chi'_-\rangle\langle \chi'_-| \rangle\nonumber\\
&&+P(A_z=0)\;\langle  |0\rangle\langle 0| \rangle+ P(A_z=1)\;\langle
 |1\rangle\langle 1| \rangle\nonumber\\
&\leq& C_{\rm LHS},
\end{eqnarray}
with $|\chi_\pm\rangle=\cos\theta|0\rangle \pm  e^{i\phi}\sin\theta|1\rangle$, $|\chi'_\pm\rangle=\cos\theta|0\rangle \mp i e^{i\phi}\sin\theta|1\rangle$, $C_{\rm LHS}={\rm Max}\{ \frac{3+\mathcal{C}_+}{2}, \;\frac{3+\mathcal{C}_-}{2}\}$,
and $\mathcal{C}_\pm=\sqrt{4\pm4\cos{2\theta}+\cos{4\theta}}$.
The equivalent 3-setting steering inequality is given by (see Appendix B)
\begin{eqnarray}\label{si-3-1 m}
 \mathcal{S}'_3(\theta,\phi)&=&\sin2\theta\cos\phi\langle A_x \sigma_x\rangle-\sin2\theta\cos\phi \langle A_y \sigma_y\rangle\nonumber\\
&  &+\sin2\theta\sin\phi\langle A_x \sigma_y\rangle+\sin2\theta\sin\phi \langle A_y \sigma_x\rangle\nonumber\\
&  &+ \langle A_z \sigma_z\rangle+2\cos2\theta\langle\sigma_z\rangle\leq C'_{\rm LHS},
\end{eqnarray}
with $C'_{\rm LHS}={\rm Max}\{\mathcal{C}_+, \mathcal{C}_-\}$. Obviously, by taking $\theta=\pi/4, \phi=0$, the inequality Eq. (\ref{si-3-1 m}) reduces to the usual 3-setting LSI in the form~\cite{saunders10}: 
\begin{eqnarray}\label{si-3-2}
\mathcal{S}'_3(\pi/4,0)&=& \langle A_x \sigma_x\rangle- \langle A_y \sigma_y\rangle
+ \langle A_z \sigma_z\rangle\leq \sqrt{3}. \label{3setting}
\end{eqnarray}

%Quantum mechanically, Alice's measurements will be represented as operators, such as
%$A_x\mapsto \sigma_x,~~A_y\mapsto\sigma_y,~~A_z\mapsto\sigma_z$. To test the GLSI

In the experiment to test the inequalities, Alice prepares two qubits and sends one of them to Bob, who trusts his own measurements but not Alice's.
Bob asks Alice to measure at random $\sigma_x$, $\sigma_y$ or $\sigma_z$ on her qubit with or simply not to perform any measurement; then Bob measures $\sigma_x$, $\sigma_y$ or $\sigma_z$ on his qubit with according to Alice's measurement.
%\textcolor{red}{Each time, Bob randomly selects one observable from $I$,$\sigma_x$, $\sigma_y$ or $\sigma_z$, and asks Alice to measure her own qubit along its corresponding direction; then Bob measures his qubit with $\sigma_x$, $\sigma_y$ or $\sigma_z$ according to the inequality settings.}
 Finally, Bob evaluates the average values $\langle\sigma_x\otimes\sigma_x\rangle$, $\langle\sigma_y\otimes\sigma_y\rangle$, $\langle\sigma_x\otimes\sigma_y\rangle$, $\langle\sigma_y\otimes\sigma_x\rangle$, $\langle\sigma_z\otimes\sigma_z\rangle$, and $\langle\openone\otimes\sigma_z\rangle$ and is therefore capable of checking whether the steering inequality Eq. (\ref{si-3-1 m}) is violated or not. In particular, for the case of pure states Eq. (\ref{decom-z}), if Alice is honest in the preparation and measurements of the states, the inequalities are violated for all values of $\alpha$ and $\varphi$ (except at $\alpha=0, \pi/2$), thereby confirming Alice's ability to steer Bob.

\begin{figure}[t]
\includegraphics[width=1\columnwidth]{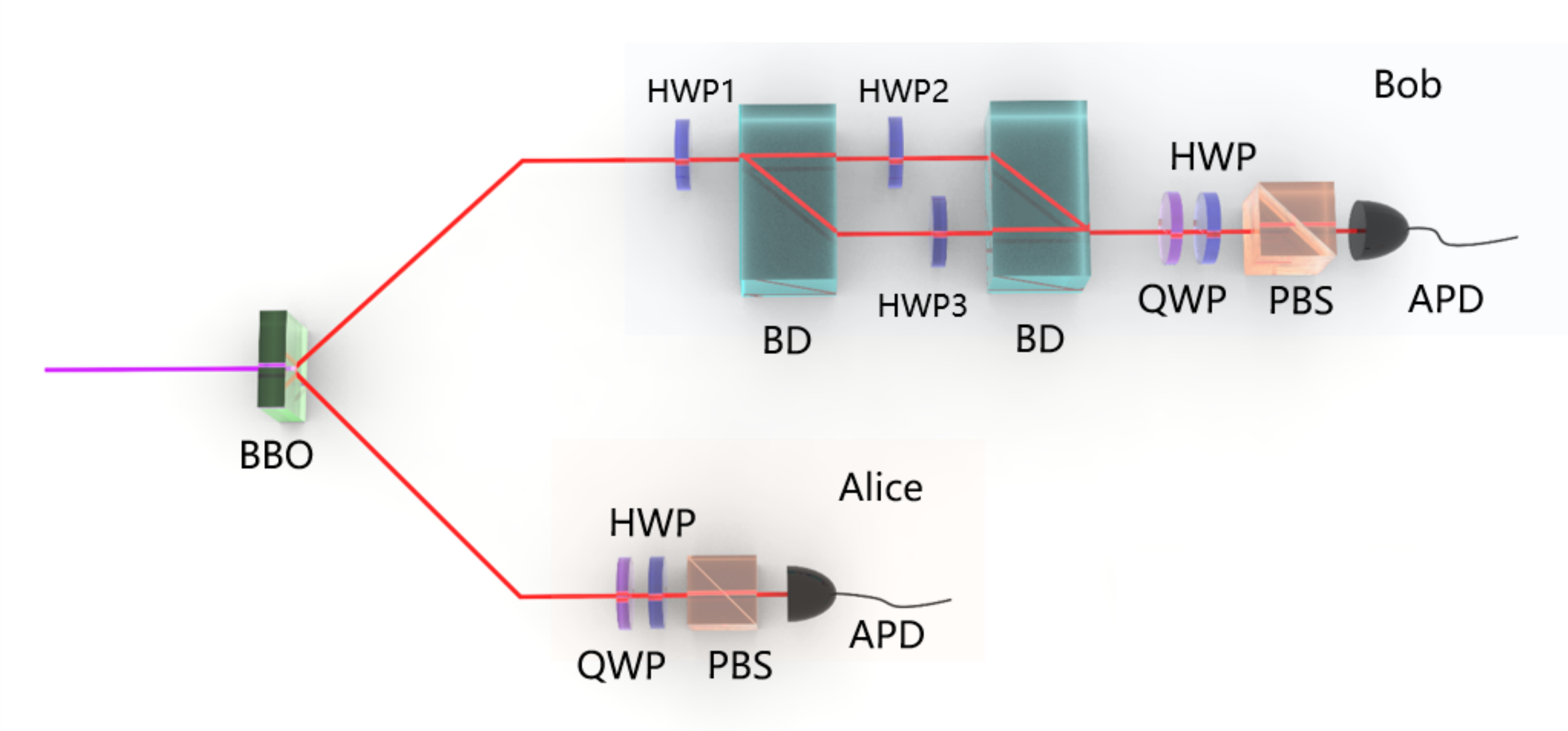}
\caption{{\bf Experimental setup}. Polarization-entangled photons pairs are generated via nonlinear crystal. An asymmetric loss interferometer along with half-wave plates (HWPs) are used to prepare two-qubit pure entangled states. The projective measurements are performed using wave plates and polarization beam splitter (PBS).}
\label{setup}
\end{figure}

\begin{figure}[t]
\includegraphics[width=0.83\columnwidth]{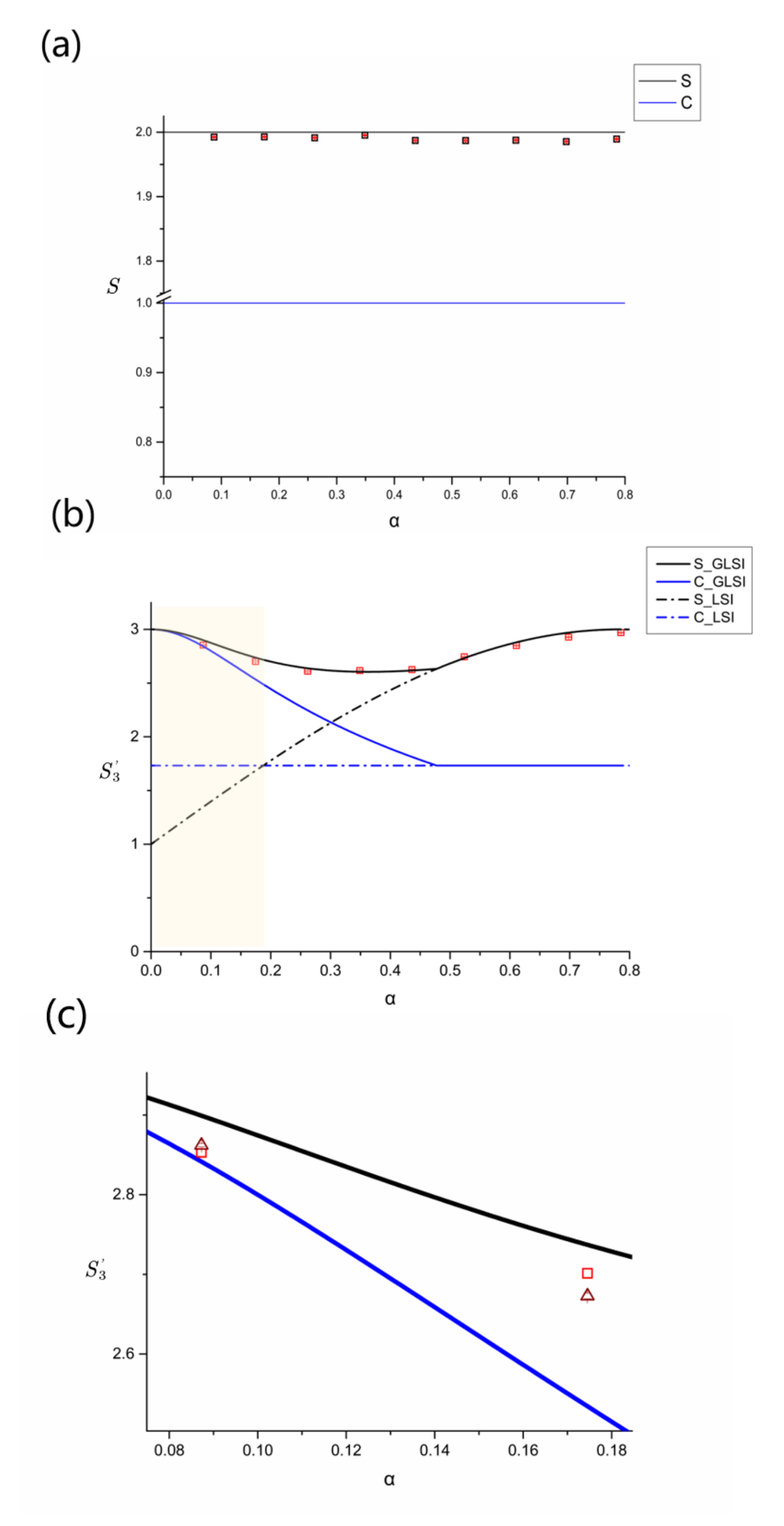}
\caption{{\bf Experimental results for pure states.} In panel (a) we show the experimental results concerning the steering paradox ``2=1''. The black and blue solid lines represent the quantum prediction $S \equiv P^{\rm QM}_{\rm total}=2$ and the classical bound $C=1$ based on the LHV models, respectively. The black cubes and the red lines show the experimental results with error bar. Panel (b) shows the experimental results for the 3-setting GLSI of Eq. (\ref{si-3-1 m}). The black and blue solid line represent the quantum and classic bound, respectively, which are obtained by maximizing the difference between  $\mathcal{S}'_3$ and  $C'_{\rm LHS}$ for any fixed $\alpha$.
The black~(blue) dot line represents the quantum violations $\langle\mathcal{S}''_3\rangle=1+2\sin(2\alpha)$~(the classical value $C=\sqrt{3}$) of the usual 3-setting LSI Eq. (\ref{si-3-2}), respectively. The red cubes are the experimental points for the inequality Eq. (\ref{si-3-1 m}). The light yellow range is $\alpha\in(0,(\arcsin\frac{\sqrt{3}-1 }{2})/2]$, where the LSI Eq. (\ref{si-3-2}) cannot detect the steerability but the GLSI can. Panel (c)  shows the experimental violation for $\alpha=\frac{\pi}{36}, \frac{\pi}{18}$. }
\label{data1}
\end{figure}

\begin{figure*}[t]
\includegraphics[width=1\textwidth]{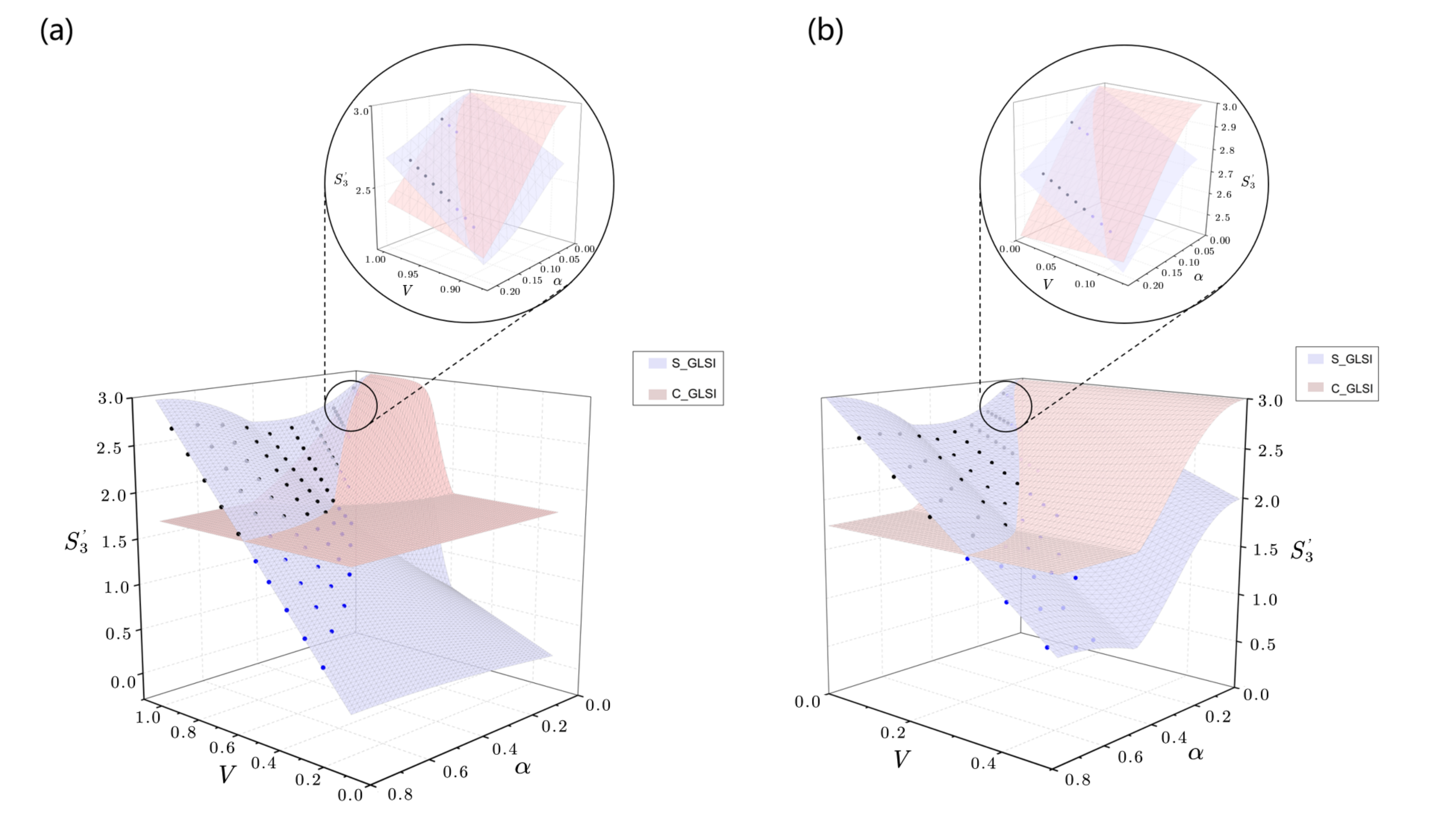}   
\caption{{\bf Experimental results for mixed states.} Panels (a) and (b) show the steering detection for the generalized Werner state $\rho_1$ and the asymmetric mixed state $\rho_2$. The light purple and pink surfaces represent the quantum value and the classical bound of the GLSI (\ref{si-3-1 m}), respectively. The black (blue) dots denote results for the quantum states that can (cannot) experimentally violate the GLSI (\ref{si-3-1 m}).  The zoom shows the area where steering cannot be detected by usual LSI (\ref{si-3-2}), whereas GLSI may be useful.}
\label{data}
\end{figure*}

\section{3. Experimental results}
%We have experimentally demonstrated the steering paradox ``$2=1$'' and quantum violations of the GLSI via the photonic system.
The experimental setup is shown in Fig.~\ref{setup}, the degenerated polarization-entangled photon pairs are created by spontaneous parametric down-conversion~\cite{Source} type-II barium borate~(BBO) crystal pumped by a 404nm laser. The initial two-photon state is singlet state $|\psi\rangle=(|HV\rangle-|VH\rangle)/\sqrt{2}$.
%To get the quantum state like (\ref{decom-z}), one should flip one of the qubits, add a relative phase between the orthogonal components and redistribute the amplitude coefficients of the entangled state. By setting HWP1 at $0^{o}$ one may switch the phase from ``minus'' to ``plus''. Then, by two half wave plate~(HWP2 and HWP3) and beam displacers~(BD), we construct an asymmetric loss interferometer. As shown in Fig.~\ref{setup} , Bob's photon passes through BD interferometer, HWP3 fixed at $45^{o}$ and  HWP2 at $\arcsin(\sqrt{\frac{1}{\sin^2\alpha}-1})\cdot\frac{90}{\pi}\in[0,45^{o}]$, which causes photon loss to adjust the amplitude and flips the qubit, to prepare desired two-qubit pure entangled state
By setting HWP1 at $0^{o}$ (correspond to a phase gate) one may switch the phase of entangled state from ``minus'' to ``plus''. HWP2, HWP3 and  beam displacers~(BD) are used to construct an asymmetric loss interferometer to adjust the amplitude and flip the qubit, where HWP3 fixed at $45^{o}$ and  HWP2 at $\arcsin(\sqrt{\frac{1}{\sin^2\alpha}-1})\cdot\frac{90}{\pi}\in[0,45^{o}]$. Therefore, we may prepare desired two-qubit state as
$ |\Psi(\alpha)\rangle= \cos\alpha |HH\rangle + \sin\alpha |VV\rangle$ (see Appendix C). Now we define $|H\rangle=|0\rangle$ and $|V\rangle=|1\rangle$, the  entangled state becomes $ |\Psi(\alpha)\rangle= \cos\alpha |00\rangle + \sin\alpha |11\rangle$. Compare with Eq. (\ref{decom-z}),
in our experiment, the phase $\varphi$ of the entangled state is set to 0. 
As illustrated in Fig.~\ref{setup}, after preparation of entangled state, we sent the first qubit to Alice, and the second to Bob. Then Alice and Bob measure their own photons through the polarization analyzer, which consists of quarter-wave plate (QWP), HWP and PBS, and test EPR steering.

First, we test the EPR paradox via steering paradox ``$2=1$'' by choosing $\alpha$ from $\frac{\pi}{36}$ to $\frac{\pi}{4}$ with an interval of $\frac{\pi}{36}$ to obtain nine different two-qubit entangled states.  In the two-setting steering scenario, Alice performs measurements on her photon along the $\hat{x}$-direction and $\hat{z}$-direction of Bloch sphere. The eigenvectors of $\sigma_{x}$ are $|\pm\rangle=(|0\rangle\pm|1\rangle)/\sqrt{2}$, which are the states on which the photon of Alice may collapse with a certain probability. The corresponding normalized conditional states for Bob are given by $|\chi_{\pm}\rangle=\cos\alpha|0\rangle\pm\sin\alpha|1\rangle$. Similarly, Bob's normalized conditional states are $|0\rangle$ and $|1\rangle$ when Alice performs corresponding measurements. As shown in Fig.~\ref{data1}(a), the experimental values $S=P^Q_{total}$ for the nine different entangled pure states largely exceed the classical prediction. The average value is $S\approx1.9899$, which is far exceeds above the classical bound predicted by LHS models. Thus, the steering paradox has been successfully demonstrated. %123��׼��%

Second, we experimentally address the violations of the GLSI using above pure states $ |\Psi(\alpha)\rangle$. We experimentally evaluate the value of  $\mathcal{S}'_3$ by using  the 3-setting steering inequality Eq. (\ref{si-3-1 m}). For simplicity, in our experiment the phase $\phi$ is set to 0, and therefore, following Eq. (\ref{si-3-1 m}), we only need to measure the following four expectation values: $\langle\sigma_{x}\otimes \sigma_{x}\rangle$, $\langle\sigma_{y}\otimes \sigma_{y}\rangle$, $\langle\sigma_{z}\otimes \sigma_{z}\rangle$,  and $\langle\sigma_{I}\otimes \sigma_{z}\rangle$. Besides, in order to experimentally observe the violation of the GLSI for any $\alpha\in(0, \pi/4]$, we have to maximize the difference between $\mathcal{S}'_3$  and classical bound $C'_{\rm LHS}$ at any fixed value $\alpha$. This is done by numerically solving the optimal solutions of $\theta$. Remarkably, for $\alpha \in(0, (\arcsin\frac{\sqrt{3}-1}{2})/2\approx \frac{\pi}{17}]$, one observes a significant violation of the inequality, which does not occur for the usual 3-setting LSI Eq. (\ref{si-3-2}). On the other hand, when $\alpha$ is close to $\pi/4$, the violation of the GLSI Eq. (\ref{si-3-1 m}) and the LSI Eq. (\ref{si-3-2}) are of the same order. The experimental results are shown in Fig.~\ref{data1}(b) and (c), which are almost indistinguishable from the theoretical predictions.

Finally, we have experimentally tested inequality Eq. (\ref{si-3-1 m}) with two types of mixed states (see Appendix C). The first one is a generalized Werner state $\rho_1$ \cite{Werner} and the second one is the asymmetric mixed state $\rho_2$ \cite{chen13}, which are given as
\begin{subequations}  \label{mixed}
\begin{eqnarray}
   \rho_{1}&=&V|\Psi(\alpha)\rangle\langle\Psi(\alpha)|+\frac{1-V}{4}\openone\otimes \openone, \label{mixed-1}\\
\rho_{2}&=&V|\Psi(\alpha)\rangle\langle\Psi(\alpha)|+(1-V)|\Phi(\alpha)\rangle\langle\Phi(\alpha)|,\label{mixed-2}
\end{eqnarray}
\end{subequations}
with $|\Phi(\alpha)\rangle=\sin\alpha|01\rangle+\cos\alpha|10\rangle$, $\alpha\in[0,\frac{\pi}{4}]$, and $V\in[0,1]$. As it is apparent from Fig.~\ref{data}(a) and Fig.~\ref{data}(b), the experimental results confirm that the GLSI has an advantage over the LSI in detecting steerability for more quantum states (One can see Appendix B for more theoretical detials).

% As it is apparent from Fig.~\ref{data}(c) and Fig.~\ref{data}(d), \textcolor{blue}{the experimental results of these two mixed states are shown, in which one can detect more steerable states in the region of $\alpha$ and $V$ by using the 3-setting GLSI, while there are no states violate the usual 3-setting LSI with the range of $\alpha\in[0,\frac{\arcsin{\frac{\sqrt{3}-1}{2}}}{2}]$ and some region (see more detail in Appendix B, especially, Fig. \ref{figwerner-1}, Fig. \ref{figwerner-2}, Fig. \ref{AVN-1} and Fig. \ref{AVN-2}).
%Our results confirm that the GLSI has an advantage over the LSI in detecting steerability for more quantum states.}

\section{4. Conclusions} In summary, we have advanced the study of  EPR paradox in two aspects: (i) We have presented a generalized steering paradox ``$k=1$'' and performed an experiment to illustrate the original EPR paradox through demonstrating the steering paradox ``$2=1$'' in a two-qubit scenario. (ii) Based on the steering paradox ``$k=1$'', we have successfully generated a $k$-setting generalized linear steering inequality, which may detect steerability of quantum states to a larger extent than the previous ones. We have also rewritten this inequality into a mathematically equivalent form, which is more suitable for experimental implementation since it allows us to measure only along the $x$-, $y$-, or $z$-axis in Bloch sphere, rather than other arbitrary directions, and thus greatly simplifying the experimental setups and improving precision. This finding is valuable for the open problem of how to optimize the measurement settings for steering verification in experiments~\cite{RMP}. Our results deepen the understanding of quantum foundations and provide an efficient way to detect the steerability of quantum states.

Recently, quantum steering has been applied to the one-side device-independent quantum key distribution protocol to secure shared keys by measuring the quantum steering inequality \cite{Branciard}. Our generalized linear steering inequality can also be applied to this scenario to implement the one-side DIQKD. In addition, our results may be applied to applications such as quantum random number generation \cite{Skrzypczyk,Guo} and quantum sub-channel discrimination \cite{Piani,Sun}.

\section{Appendix A. Generalized linear steering inequality obtained from the general steering paradox ``$k=1$''.} %Bell's inequalities can be naturally derived from the GHZ and the Hardy paradoxes. This is also the case for the steering paradox. 
Actually, from the steering paradox ``$k=1$'', one can naturally derive a $k$-setting generalized linear steering inequality (GLSI), which include the usual LSI~\cite{saunders10} as special case.

The derivation procedure is as follows: In the steering scenario $\{\hat{n}_1, \hat{n}_2, \cdots, \hat{n}_k \}$, Alice performs $k$ projective measurements $\hat{P}_a^{\hat{n}_j}$. %along $\hat{n}_j$-directions with measurement outcomes $a=0$ and 1.
 For the pure state $ |\Psi(\theta,\phi)\rangle$ as shown in Eq. (\ref{decom-z}), it can have an equivalent but more general decomposition along the $\hat{n}$-direction as Eq. (\ref{decom-n}) .
%\begin{eqnarray}\label{decom-n}
% |\Psi(\theta,\phi)\rangle &=&\cos\theta|00\rangle+e^{i\phi} \sin\theta|11\rangle\nonumber\\
% &=& |+\hat{n}\rangle |\chi_{+\hat{n}}\rangle+ |-\hat{n}\rangle |\chi_{-\hat{n}}\rangle,
% \end{eqnarray}
%where $|+ \hat{n}\rangle$ and  $|- \hat{n}\rangle$ are the eigenstates of $\hat{n} \cdot \vec{\sigma}$ with eigenvalues are $+1$ and $-1$ respectively, and
%\begin{eqnarray}\label{chiold}
%  |\chi_{+\hat{n}}\rangle= \langle +\hat{n}|\Psi(\theta,\phi)\rangle, \;\;\;   |\chi_{-\hat{n}}\rangle= \langle -\hat{n}|\Psi(\theta,\phi)\rangle.
% \end{eqnarray}
%are the collapsed pure states (unnormalized) for Bob's qubit. 
Explicitly, Alice's projective measurements could be rewritten as
\begin{subequations}  \label{proj}
\begin{eqnarray}
  \hat {P}^{\hat{n}_j}_0&=&\frac{\openone+ {\hat{n}_j}\cdot {\vec \sigma}}{2}=|+ \hat{n}\rangle \langle +\hat{n}|,\\
  \hat {P}^{\hat{n}_j}_1&=&\frac{\openone- {\hat{n}_j}\cdot {\vec \sigma}}{2}=|- \hat{n}\rangle \langle -\hat{n}|.
\end{eqnarray}
\end{subequations}
Based on the two-qubit pure state $|\Psi (\theta,\phi)\rangle$, for the $j$-th projective measurement $\hat{P}_a^{\hat{n}_j}$ of Alice, Bob will have the unnormalized conditional states as
\begin{subequations}
\begin{eqnarray}
\tilde{\rho}^{\hat{n}_j}_0={\rm tr}_A[(\hat{P}_0^{\hat{n}_j}\otimes \openone)\;|\Psi\rangle\langle \Psi|]=|\chi_{+\hat{n}_j}\rangle\langle \chi_{+\hat{n}_j}|,\\
\tilde{\rho}^{\hat{n}_j}_1={\rm tr}_A[(\hat{P}_1^{\hat{n}_j}\otimes \openone)\;|\Psi\rangle\langle \Psi|]=|\chi_{-\hat{n}_j}\rangle\langle \chi_{-\hat{n}_j}|,
\end{eqnarray}
\end{subequations}
and from which one has the normalized conditional states as
\begin{subequations}
\begin{eqnarray}
{\rho}^{\hat{n}_j}_0&=&\frac{\tilde{\rho}^{\hat{n}_j}_0}{{\rm
tr}(\tilde{\rho}^{\hat{n}_j}_0)}=|\chi_{+}^j\rangle\langle \chi_{+}^j|,\\
{\rho}^{\hat{n}_j}_1&=&\frac{\tilde{\rho}^{\hat{n}_j}_1}{{\rm
tr}(\tilde{\rho}^{\hat{n}_j}_1)}=|\chi_{-}^j\rangle\langle \chi_{-}^j|,
\end{eqnarray}
\end{subequations}
with
\begin{eqnarray}
|\chi_{+}^j\rangle=\frac{|\chi_{+\hat{n}_j}\rangle}{\sqrt{{\rm tr}[|\chi_{+\hat{n}_j}\rangle\langle \chi_{+\hat{n}_j}|]}},\\
\;\;\;
|\chi_{-}^j\rangle=\frac{|\chi_{-\hat{n}_j}\rangle}{\sqrt{{\rm tr}[|\chi_{-\hat{n}_j}\rangle\langle \chi_{-\hat{n}_j}|]}},
\end{eqnarray}
are pure states. Obviously, for the pure state $|\Psi\rangle$, the following probability relation is always hold:
%\begin{widetext}
\begin{eqnarray}\label{prob}
{\rm tr}[(\hat{P}_0^{\hat{n}_j}\otimes |\chi_{+}^j\rangle\langle \chi_{+}^j| )\;|\Psi\rangle\langle \Psi|]+\\
{\rm tr}[(\hat{P}_1^{\hat{n}_j}\otimes |\chi_{-}^j\rangle\langle \chi_{-}^j| )\;|\Psi\rangle\langle \Psi|]=\\
{\rm tr}[|\chi_{+\hat{n}_j}\rangle\langle \chi_{+\hat{n}_j}|]+{\rm tr}[|\chi_{-\hat{n}_j}\rangle\langle \chi_{-\hat{n}_j}|] \equiv 1.
\end{eqnarray}
%\end{widetext}

The quantity in the left-hand-side of Eq. (\ref{prob}) can be used to construct the steering inequality, where for the Alice's side we need to replace the quantum measurement operator $\hat{P}_a^{\hat{n}_j}$ by their correspondingly classical probabilities $P(A_{n_j}=a)$. Then we immediately have the $k$-setting GLSI as
\begin{eqnarray}\label{k-setting inequality}
\mathcal{S}_k&=& \sum_{j=1}^k \biggr(\sum_{a=0}^1 P(A_{n_j}=a)\;\langle  {\rho}^{\hat{n}_j}_a \rangle \biggr)\leq C_{\rm LHS},
\end{eqnarray}
where $P(A_j=a)$ is the classical probability of the $j$-th measurement of Alice with outcome $a$,
\begin{eqnarray}\label{projpm}
{\rho}^{\hat{n}_j}_0=|\chi^j_+\rangle\langle \chi^j_+|,\;\;\; {\rho}^{\hat{n}_j}_1=|\chi^j_-\rangle\langle \chi^j_-|,
\end{eqnarray}
 denote the projective measurements in Bob's side, $C_{\rm LHS}$ is the classical bound that determined by the maximal eigenvalue of the steering parameter $\mathcal{S}_k$.  By definition, it is easy to verify directly that for the pure state (\ref{decom-z}) the quantum prediction of $\mathcal{S}_k$ is equal to $k$.

\emph{Remark 1.---}In Ref. \cite{saunders10}, the usual $k$-setting linear steering inequality is given as
\begin{eqnarray}\label{kLSI}
\mathcal{S}^{\rm LSI}_k&=& \sum_{j=1}^k  A_j \langle  \hat{m}_j  \cdot \vec{\sigma} \rangle \leq C^{\rm LSI}_{\rm LHS},
\end{eqnarray}
here $C^{\rm LSI}_{\rm LHS}$ denotes the classical bound for the LSI.
Note that for the LSI Eq. (\ref{kLSI}), quantum mechanically Alice will perform $k$ measurements (corresponds to $\hat{A}_j=\hat{n}_j\cdot \vec{\sigma}$), and Bob will also perform $k$ measurements (corresponds to $\hat{B}_j=\hat{m}_j\cdot \vec{\sigma}$). In the following we would like to show that the LSI is a special case of the GLSI as given in Eq. (\ref{k-setting inequality}).

Let us rewrite the projective measurements Eq. (\ref{projpm}) in the Bloch-representation as
\begin{eqnarray}\label{projpm2}
{\rho}^{\hat{n}_j}_0&=&|\chi^j_+\rangle\langle \chi^j_+|=\frac{1}{2}(\openone + \hat{m}^j_+ \cdot \vec{\sigma}),\nonumber\\
 {\rho}^{\hat{n}_j}_1&=&|\chi^j_-\rangle\langle \chi^j_-|=\frac{1}{2}(\openone + \hat{m}^j_- \cdot \vec{\sigma}),
\end{eqnarray}
and we denote
\begin{eqnarray}\label{classp}
P(A_j=0)=\frac{1+A_j}{2}, \;\;\; P(A_j=1)=\frac{1-A_j}{2},
\end{eqnarray}
then by substituting Eqs. (\ref{projpm2}) and (\ref{classp}) into the inequality Eq.  (\ref{k-setting inequality}), we have
%\begin{widetext}
\begin{eqnarray}\label{k-setting 2}
\mathcal{S}_k&=& \sum_{j=1}^k \biggr(\frac{1+A_j}{2} \;\langle \frac{1}{2}(\openone + \hat{m}^j_+ \cdot \vec{\sigma}) \rangle + \\
&&\frac{1-A_j}{2} \;\langle \frac{1}{2}(\openone + \hat{m}^j_- \cdot \vec{\sigma}) \rangle\biggr)\leq C_{\rm LHS}.
\end{eqnarray}
%\end{widetext}
Note that for the GLSI Eq. (\ref{k-setting 2}), quantum mechanically Alice will perform $k$ measurements (corresponds to $\hat{A}_j=\hat{n}_j\cdot \vec{\sigma}$), and Bob will perform $2k$ measurements
(corresponds to $\hat{B}_j^+=\hat{m}^j_+\cdot \vec{\sigma}$ and $\hat{B}_j^-=\hat{m}^j_-\cdot \vec{\sigma}$, if $\hat{m}^j_+ \neq \pm\hat{m}^j_-$). In the following we would like to show that the LSI is a special case of the GLSI as given in Eq. (\ref{k-setting inequality}).

Let
\begin{eqnarray}\label{nj}
\hat{n}_j=(\sin\tau \cos{\gamma}, \sin\tau \sin{\gamma}, \cos\tau),
\end{eqnarray}
then
\begin{eqnarray}\label{pmnj}
|+\hat{n}_j\rangle=\left(
                     \begin{array}{l}
                       \cos\frac{\tau}{2} \\
                       \sin\frac{\tau}{2} e^{i \gamma} \\
                     \end{array}
                   \right),\;\;\;
|-\hat{n}_j\rangle=\left(
                     \begin{array}{l}
                       \sin\frac{\tau}{2} \\
                       -\cos\frac{\tau}{2} e^{i \gamma} \\
                     \end{array}
                   \right),
\end{eqnarray}
which are eigenstates of $\hat{n}_j\cdot \vec{\sigma}$. From Eq. (\ref{decom-n}), one can explicitly have

\begin{eqnarray}\label{chinew}
  |\chi_{+\hat{n}_j}\rangle &=& \langle +\hat{n}_j|\Psi(\theta,\phi)\rangle \\
 & =&\cos\frac{\tau}{2}\cos\theta|0\rangle+e^{i(\phi-\gamma)} \sin\frac{\tau}{2}\sin\theta|1\rangle, \nonumber\\
  |\chi_{-\hat{n}_j}\rangle &=& \langle -\hat{n}_j|\Psi(\theta,\phi)\rangle \\
 & =&\sin\frac{\tau}{2}\cos\theta|0\rangle-e^{i(\phi-\gamma)} \cos\frac{\tau}{2}\sin\theta|1\rangle, \end{eqnarray}
 
which yields
\begin{eqnarray}\label{chinew1}
 \langle \chi_{-\hat{n}_j} |\chi_{+\hat{n}_j}\rangle &=& \cos\frac{\tau}{2} \sin\frac{\tau}{2} (\cos^2\theta-\sin^2\theta).
  \end{eqnarray}
Namely, if $\theta=\pi/4$, the states $|\chi_{+\hat{n}_j}\rangle$ and $|\chi_{-\hat{n}_j}\rangle$ are orthogonal, then from Eq. (\ref{projpm2}) one immediately knows that the two Bloch vectors are antiparallel, i.e.,
\begin{eqnarray}\label{anti}
 \hat{m}^j_+ =- \hat{m}^j_- \equiv \hat{m}^j.
\end{eqnarray}
Substituting the relation Eq. (\ref{anti}) into the inequality Eq. (\ref{k-setting 2}), we have
\begin{equation}\label{k-setting 3a}
\mathcal{S}_k= \frac{1}{2} \sum_{j=1}^k \biggr(1 +\; A_j \langle  \hat{m}^j \cdot \vec{\sigma} \rangle \biggr) \nonumber \leq C_{\rm LHS}
\end{equation}

i.e.,

\begin{eqnarray}\label{kLSI-2}
\mathcal{S}^{\rm LSI}_k&=& \sum_{j=1}^k  A_j \langle  \hat{m}_j  \cdot \vec{\sigma} \rangle \leq 2\; C_{\rm LHS}-k.
\end{eqnarray}
 proving the usual LSI is a special case of the GLSI. In short, for an arbitrary two-qubit pure entangled state $|\Psi(\theta,\phi)\rangle$, one can have a steering paradox ``$k=1$''~\cite{chen16}, based on which one can derive a general linear steering inequality as shown in Eq. (\ref{k-setting inequality}). If one further fixes the  parameter $\theta$ as  $\theta=\pi/4$ (in this case $|\Psi(\theta=\pi/4,\phi)\rangle$ is the maximally entangled state), then the GLSI reduces to the usual LSI.

\emph{Remark 2.---}We may rewrite the GLSI Eq. (\ref{k-setting 2}) in a mathematically equivalent form, but which is more friendly for experimental implements. Let us denote
\begin{eqnarray}\label{mm}
\hat{m}^j_+=(m^j_{+x}, m^j_{+y},m^j_{+z}), \;\;\; \hat{m}^j_-=(m^j_{-x}, m^j_{-y},m^j_{-z}),
\end{eqnarray}
then from the inequality Eq. (\ref{k-setting 2}) one has
\begin{widetext}
\begin{eqnarray}\label{k-setting 3}
\mathcal{S}_k&=& \sum_{j=1}^k \biggr(\frac{1+A_j}{2} \;\langle \frac{1}{2}(\openone + \hat{m}^j_+ \cdot \vec{\sigma}) \rangle + \frac{1-A_j}{2} \;\langle \frac{1}{2}(\openone + \hat{m}^j_- \cdot \vec{\sigma}) \rangle\biggr)\nonumber\\
&=& \sum_{j=1}^k \biggr(\frac{1}{2}+\frac{1+A_j}{4} \;(\hat{m}^j_{+x}\langle  \vec{\sigma}_x \rangle+\hat{m}^j_{+y}\langle  \vec{\sigma}_y \rangle+\hat{m}^j_{+z}\langle  \vec{\sigma}_z \rangle) + \frac{1-A_j}{4} \;\;(\hat{m}^j_{-x}\langle  \vec{\sigma}_x \rangle+\hat{m}^j_{-y}\langle  \vec{\sigma}_y \rangle+\hat{m}^j_{-z}\langle  \vec{\sigma}_z \rangle)\biggr)\nonumber\\
&\leq& C_{\rm LHS}.
\end{eqnarray}
\end{widetext}
The remarkable point for the inequality Eq. (\ref{k-setting 3}) is that Bob always measures his particle in three directions of $x, y, z$, which not only greatly reduces the numbers of measurements and but also need not tune the measurement direction to other directions.

\section{Appendix B. EPR steering by using the 3-setting generalized LSI.}
In this experimental work, we shall demonstrate EPR steering for the two-qubit generalized Werner state  by using the generalized linear steering inequality. We focus on the 3-setting  GLSI. In the steering scenario $\{\hat{x}, \hat{y}, \hat{z} \}$, Alice performs projective measurements on her qubit along the $\hat{x}$-, $\hat{y}$- and $\hat{z}$-directions, from Eq. (\ref{k-setting inequality}) one immediately has
\begin{eqnarray}\label{3-setting inequality}
\mathcal{S}_3&=&P(A_x=0)\; \langle
 |\chi_+\rangle\langle \chi_+|
\rangle+P(A_x=1)\;\langle |\chi_-\rangle\langle \chi_-| \rangle
\nonumber\\
&&+ P(A_y=0)\; \langle
 |\chi'_+\rangle\langle \chi'_+|
\rangle+P(A_y=1)\;\langle |\chi'_-\rangle\langle \chi'_-| \rangle\nonumber\\
&&+P(A_z=0)\;\langle  |0\rangle\langle 0| \rangle+ P(A_z=1)\;\langle
 |1\rangle\langle 1| \rangle \nonumber\\
&\leq& C_{\rm LHS},
\end{eqnarray}
with
\begin{eqnarray}
|\chi_\pm\rangle & = & \cos\theta|0\rangle \pm  e^{i\phi}\sin\theta|1\rangle,\nonumber\\
|\chi'_\pm\rangle& = &\cos\theta|0\rangle \mp i e^{i\phi}\sin\theta|1\rangle.
\end{eqnarray}
One can have
\begin{eqnarray}\label{allchi}
|\chi_+\rangle \langle \chi_+|& = & \frac{1}{2}(\openone+\hat{m}_+\cdot \vec{\sigma}),\;\;\;
|\chi_-\rangle \langle \chi_-|=\frac{1}{2}(\openone+\hat{m}_-\cdot \vec{\sigma}),\nonumber\\
|\chi'_+\rangle \langle \chi'_+|& = & \frac{1}{2}(\openone+\hat{m}'_+\cdot \vec{\sigma}), \;\;\;
|\chi'_-\rangle \langle \chi'_-| =  \frac{1}{2}(\openone+\hat{m}'_-\cdot \vec{\sigma}),\nonumber\\
|0\rangle\langle 0| & = & \frac{1}{2}(\openone + \sigma_z), \;\;\;
|1\rangle\langle 1| =  \frac{1}{2}(\openone - \sigma_z),
\end{eqnarray}
with

\begin{eqnarray}\label{allm}
 \hat{m}_+ &=&(\sin 2\theta \cos\phi,\sin 2\theta \sin\phi,\cos 2\theta ),\nonumber\\
 \hat{m}_- %&=&(\sin (-2\theta) \cos\phi,\sin (-2\theta) \sin\phi,\cos (-2\theta) )\\
 &=&(-\sin 2\theta \cos\phi,-\sin 2\theta \sin\phi,\cos 2\theta ),\nonumber\\
 \hat{m}'_+%&=&(\sin 2\theta \cos(\phi-\frac{\pi}{2}),\sin 2\theta \sin(\phi-\frac{\pi}{2}),\\
% &&\cos 2\theta )
 &=&(\sin 2\theta \sin\phi,-\sin 2\theta \cos\phi,\cos 2\theta ),\nonumber\\
 \hat{m}'_-%&=&(\sin 2\theta \cos(\phi+\frac{\pi}{2}),\sin 2\theta \sin(\phi+\frac{\pi}{2}),\\&&\cos 2\theta )
 &=&(-\sin 2\theta \sin\phi,\sin 2\theta \cos\phi,\cos 2\theta ).
\end{eqnarray}
We now come to compute the the classical bound $C_{\rm LHS}$. Because $P(A_i=0)+P(A_i=1)=1$,  ($i=x, y, z$),  i.e., $P(A_i=0)$ is exclusive with
$P(A_i=1)$, if $P(A_i=0)=1$, then one must have $P(A_i=1)=0$. For the inequality (Eq. \ref{3-setting inequality}), there
are totally eight combinations:

(i) $P(A_z=0)=P(A_x=0)=P(A_y=0)=1$, then the left-hand side of the inequality Eq. 
(\ref{3-setting inequality}) is
\begin{eqnarray}
  \biggr\langle
 |\chi_+\rangle\langle \chi_+|+
 |\chi'_+\rangle\langle \chi'_+| +  |0\rangle\langle 0|
\biggr\rangle,
\end{eqnarray}
for such matrix, its two eigenvalues are

\begin{eqnarray}
\frac{3+\sqrt{4-4\cos{2\theta}+\cos{4\theta}}}{2},\\
 \;\;\; \frac{3-\sqrt{4-4\cos{2\theta}+\cos{4\theta}}}{2}.
\end{eqnarray}

Similarly, for $P(A_z=0)=P(A_x=0)=P(A_y=1)=1$, $P(A_z=0)=P(A_x=1)=P(A_y=0)=1$, $P(A_z=0)=P(A_x=1)=P(A_y=1)=1$.

(ii) $P(A_z=1)=P(A_x=0)=P(A_y=0)=1$, then the left-hand side of the inequality Eq. 
(\ref{3-setting inequality}) is
\begin{eqnarray}
  \biggr\langle
 |\chi_+\rangle\langle \chi_+|+
 |\chi'_+\rangle\langle \chi'_+| +  |1\rangle\langle 1|
\biggr\rangle,
\end{eqnarray}
for such matrix, its two eigenvalues are

\begin{eqnarray}
\frac{3+\sqrt{4+4\cos{2\theta}+\cos{4\theta}}}{2},\\
 \;\;\; \frac{3-\sqrt{4+4\cos{2\theta}+\cos{4\theta}}}{2}.
\end{eqnarray}

Similarly, for $P(A_z=1)=P(A_x=0)=P(A_y=1)=1$, $P(A_z=1)=P(A_x=1)=P(A_y=0)=1$, $P(A_z=1)=P(A_x=1)=P(A_y=1)=1$.

Thus, in summary, the classical bound is given by (here $\theta\in
(0, \pi/2)$)
\begin{eqnarray}\label{bound2}
C_{\rm LHS}={\rm Max}\{ \frac{3+\mathcal{C}_+}{2}, \;\frac{3+\mathcal{C}_-}{2}\},
\end{eqnarray}
with
\begin{eqnarray}\label{bound3}
\mathcal{C}_\pm=\sqrt{4\pm4\cos{2\theta}+\cos{4\theta}}.
\end{eqnarray}
Obviously the classical bound $C_{\rm LHS}\leq 3$, however, for any two-qubit pure entangled state $|\Psi(\theta,\phi)\rangle$ one has $\mathcal{S}^{QM}_3=3$, thus
any two-qubit pure entangled state violates the steering inequality Eq. (\ref{3-setting inequality}).

Let
\begin{equation}\label{pxyz}
P(A_i=0)=\frac{1+A_i}{2}, \;\;\; P(A_i=1)=\frac{1-A_i}{2}, 
\end{equation}
where ($i=x, y, z$. Substituting Eqs. (\ref{allchi}), (\ref{allm}) and  (\ref{pxyz}) into the inequality Eq. (\ref{3-setting inequality}) and after simplify, one  obtains  the equivalent 3-setting steering inequality as
\begin{eqnarray}\label{si-3-1}
 \mathcal{S}'_3&=&\sin2\theta\cos\phi\langle A_x \sigma_x\rangle-\sin2\theta\cos\phi \langle A_y \sigma_y\rangle\nonumber\\
&  &+\sin2\theta\sin\phi\langle A_x \sigma_y\rangle+\sin2\theta\sin\phi \langle A_y \sigma_x\rangle\nonumber\\
&  &+ \langle A_z \sigma_z\rangle+2\cos2\theta\langle\sigma_z\rangle\leq C'_{\rm LHS},
\end{eqnarray}
with $C'_{\rm LHS}={\rm Max}\{\mathcal{C}_+, \mathcal{C}_-\}$. Obviously, by taking $\theta=\pi/4, \phi=0$, the inequality Eq. (\ref{si-3-1}) reduces to  the usual 3-setting LSI (an equivalent form) in~\cite{saunders10}.

\begin{figure}[ht]
\begin{center}
\includegraphics[width=0.4\textwidth]{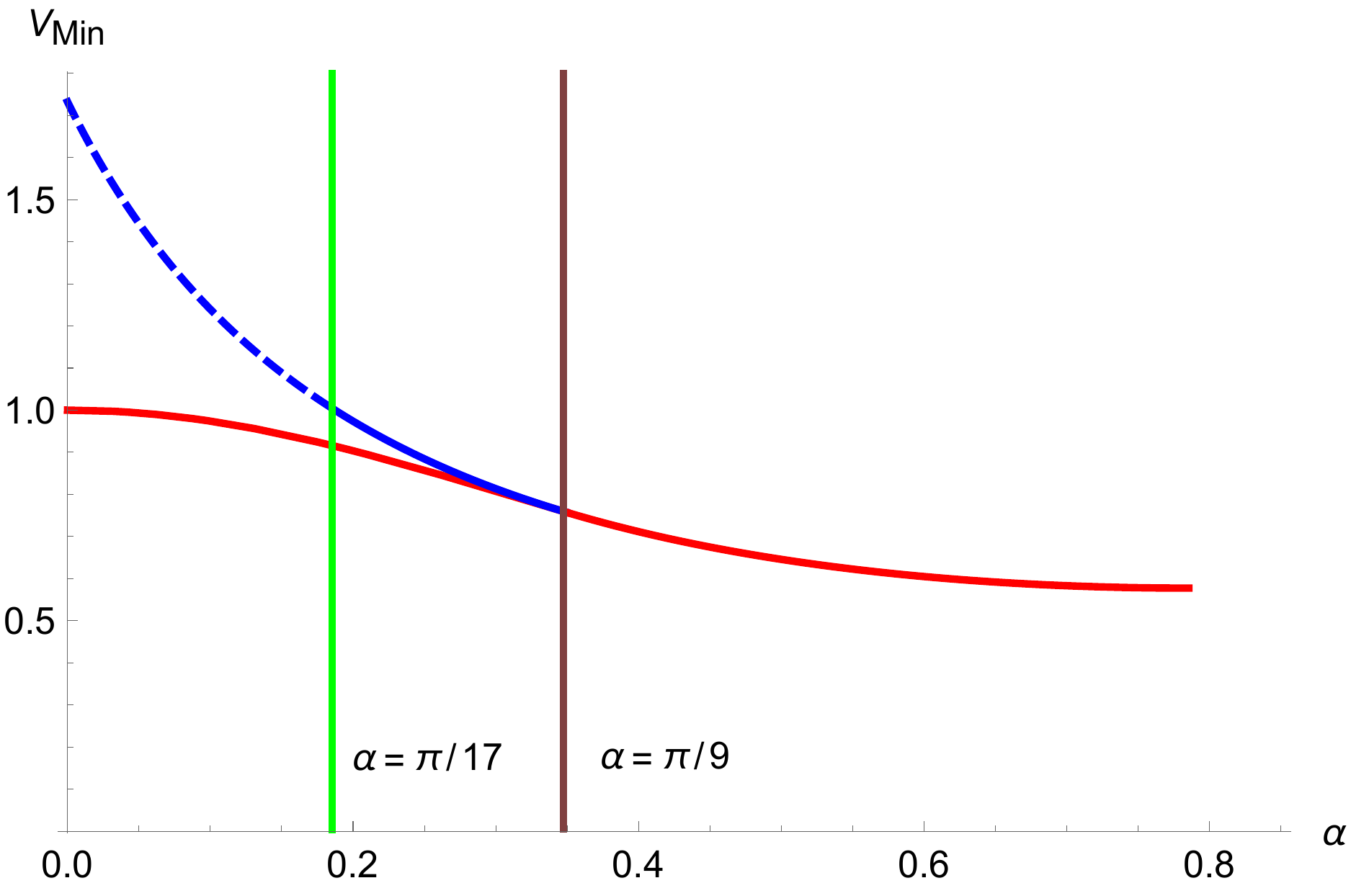}
\vspace{-3 pt}\caption{\label{figwerner-1} Detecting EPR steerability of the generalized Werner state by using the usual 3-setting LSI (Blue line) and the 3-setting GLSI (Red line). For a fixed parameter $\alpha$, the threshold value of the visibility is given by $V_{\mathrm{min}}$, below which the steering inequalities cannot be violated. It can be observed that the GLSI is stronger than the usual LSI in detecting the EPR steerability.}
\end{center}
\end{figure}

\begin{figure}[ht]
\begin{center}
\includegraphics[width=0.423\textwidth]{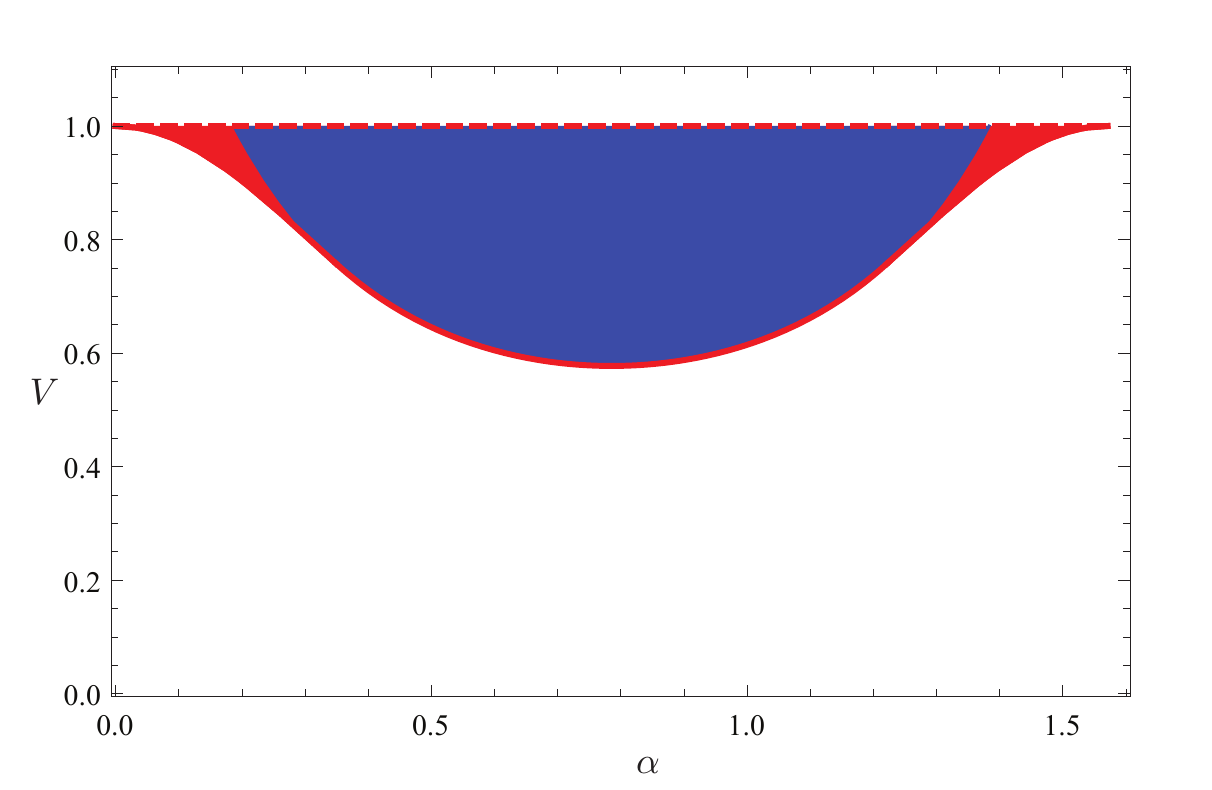}
\vspace{-10 pt}\caption{\label{figwerner-2} The generalized Werner states violate the usual 3-setting LSI in the blue region, and violate the 3-setting generalized LSI in the red region. It can be observed that the GLSI is stronger than the usual LSI in detecting the EPR steerability.}
\end{center}
\end{figure}

%\emph{Remark 4.---}The usual 3-setting linear steering inequality is given by ~\cite{saunders10}.
%\begin{eqnarray}\label{usualL}
%\mathcal{S}'_3=A_1 \langle \sigma_x\rangle + A_2 \langle \sigma_y\rangle+A_3 \langle \sigma_z\rangle\leq \sqrt{3}.
%\end{eqnarray}
%The generalized 3-setting LSI is stronger than the usual 3-setting LSI in detecting EPR steerability. In the following we would like to provide three examples.

\begin{figure}
[ht]
\begin{center}
\includegraphics[width=0.4\textwidth]{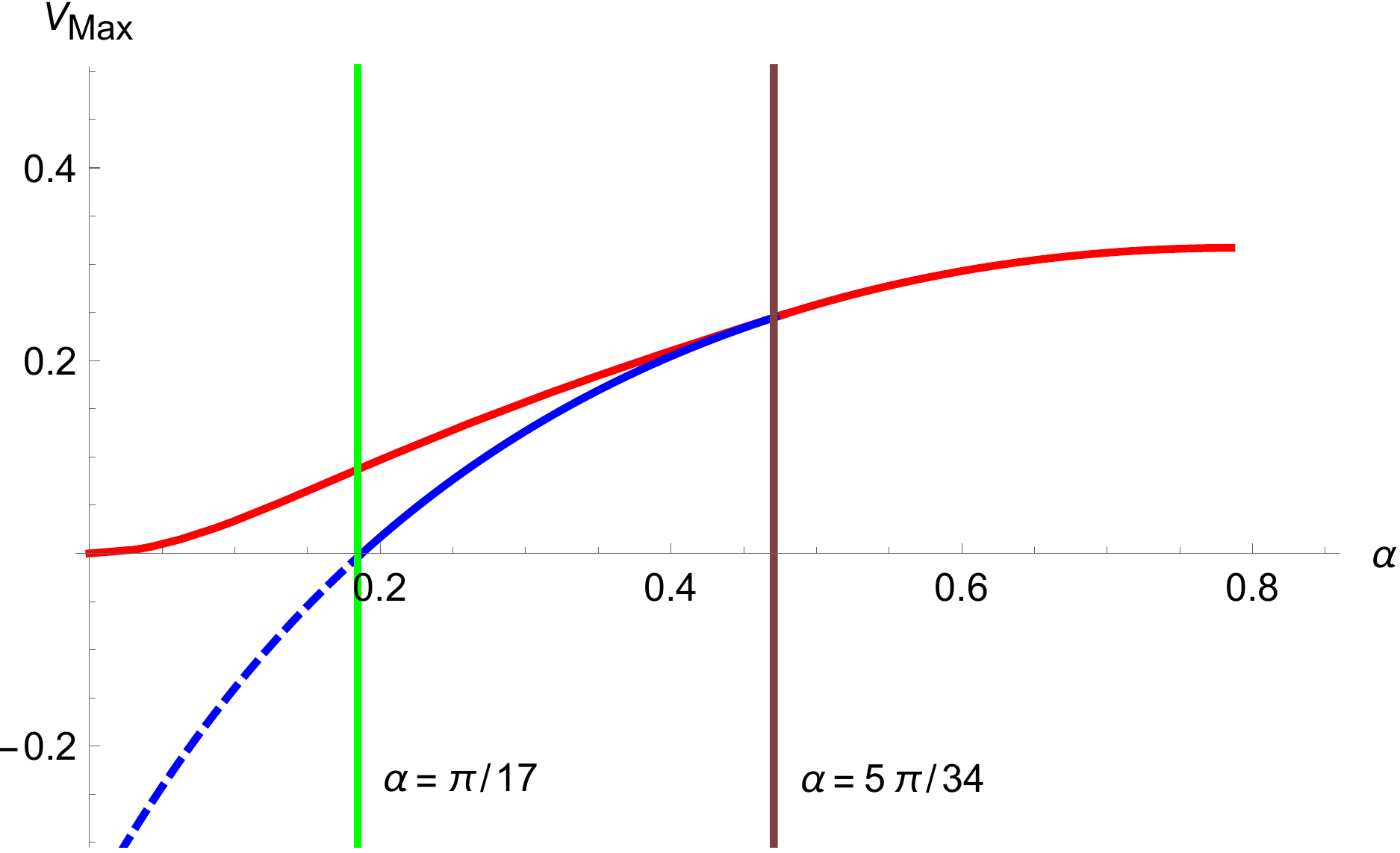}
\vspace{0 pt}\caption{\label{AVN-1} Detecting EPR steerability of the mixed state Eq. (\ref{avnstate}) by using the usual 3-setting LSI (Blue line) and the 3-setting GLSI (Red line). For a fixed parameter $\alpha$, the threshold value of the visibility is given by $V_{\mathrm{max}}$, above which the steering inequalities cannot be violated. It can be observed that the GLSI is stronger than the usual LSI in detecting the EPR steerability.}%
\end{center}
\end{figure}

\begin{figure}
[ht]
\begin{center}
\includegraphics[width=0.42\textwidth]{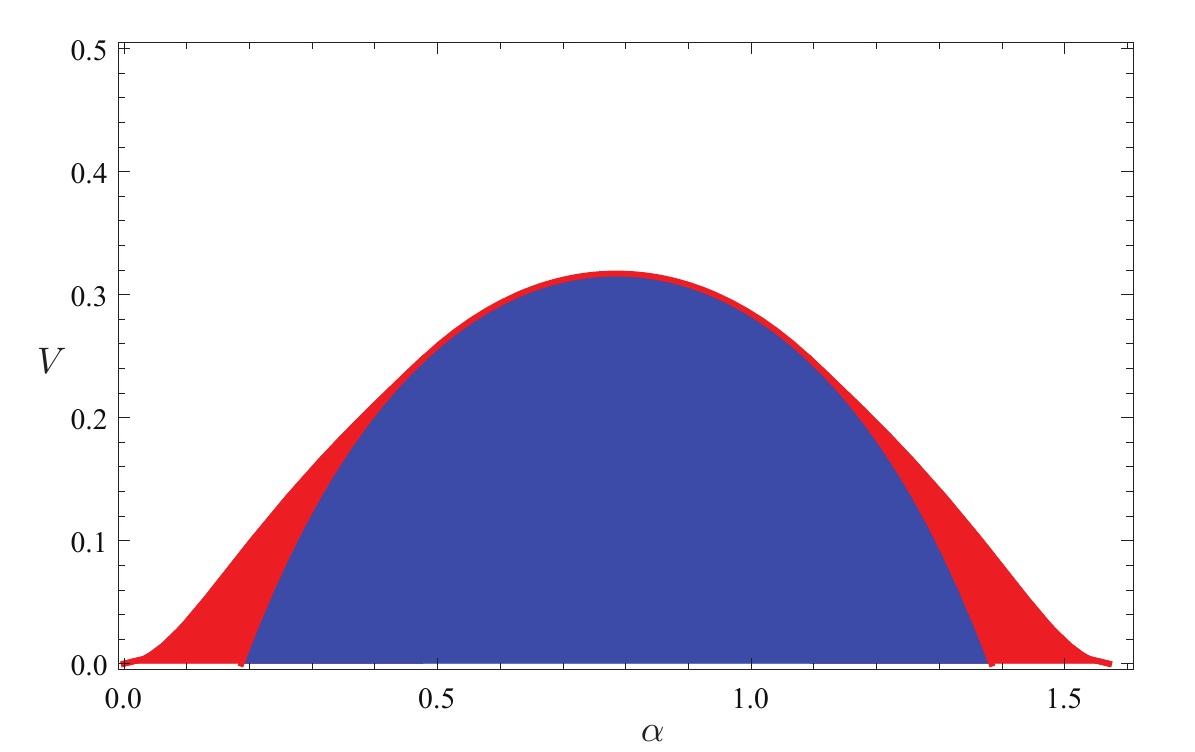} 
\vspace{-10 pt}\caption{\label{AVN-2} The mixed states Eq. (\ref{avnstate}) violate the usual 3-setting LSI in the blue region, and violate the 3-setting GLSI in the red region. It can be observed that the GLSI is stronger than the usual LSI in detecting the EPR steerability.}%
\end{center}
\end{figure}

\emph{Example 1.---} Let us consider the two-qubit pure state
$|\Psi(\alpha)\rangle=\cos\alpha|00\rangle+\sin\alpha|11\rangle$,
its maximal violation value for the usual 3-setting steering inequality Eq. (\ref{si-3-2}) is
 $1+2\sin{2\alpha}$. Hence, only when $\alpha>\frac{\arcsin{\frac{\sqrt{3}-1}{2}}}{2}\approx 0.1873$, the usual linear steering inequality can be violated. However, the pure state violates the GLSI Eq. (\ref{3-setting inequality}) for the whole region $\alpha\in (0, \pi/2)$, thus the generalized LSI is stronger than the usual LSI in detecting the EPR steerability of pure entangled states.

% In our experimental work, we will compare the performance between the 3-setting generalized LSI and the usual 3-setting steering inequality for some mixed states (which include the pure entangled state as a special case) in detail. See the following two examples.

 \emph{Example 2.---}Let us consider the two-qubit generalized Werner state
\begin{eqnarray}\label{wer}
\rho_1=\rho_{AB}(\alpha, V)= V |\Psi(\alpha)\rangle \langle
\Psi(\alpha)|+\frac{1-V}{4} \openone\otimes\openone,
\end{eqnarray}
with $\alpha\in[0,\frac{\pi}{4}], V\in[0,1]$.

For the state Eq. (\ref{wer}), we come to compare the performance between the usual 3-setting LSI (Blue line) and the 3-setting GLSI (Red line) in Fig. \ref{figwerner-1}. For the usual LSI, the threshold value of the visibility is given by $V_{\mathrm{Min}}=\frac{\sqrt{3}}{1+2\sin{2\alpha}}\approx 0.1873$, below which the usual LSI cannot be violated. Namely, it can be concluded that there are no states violate the usual 3-setting LSI with the range of $\alpha\in[0,\frac{\arcsin{\frac{\sqrt{3}-1}{2}}}{2}]$. However, the 3-setting GLSI can detect  more steerable states in the region of $\alpha$ and $V$, which can be calculated numerically. See Fig.~\ref{figwerner-1} and Fig.~\ref{figwerner-2}, for example, when $\alpha=\frac{\arcsin{\frac{\sqrt{3}-1}{2}}}{2} (\approx\frac{\pi}{17})$, there are no states violate the usual 3-setting LSI, however, the GLSI still can detect quantum states in the region $V\in [V_{\mathrm{Min}}\approx0.914, 1]$. In the range of $\alpha\in[\frac{\arcsin{\frac{\sqrt{3}-1}{2}}}{2}$, $\alpha_{B}]$,$\alpha_{B}\approx 0.3508\approx\frac{\pi}{9}$, there exist states violate the usual LSI, but the corresponding lower bound $V_{\mathrm{Min}}$ is larger than that of the 3-setting generalized LSI. In the range of $\alpha\in[\alpha_{B},\frac{\pi}{4}]$, Both inequalities are almost equivalent in the task of steering test. Especially, when $\alpha=\frac{\pi}{4}$, both of them achieve $V_{\mathrm{Min}}=\frac{\sqrt{3}}{3}$.

 \emph{Example 3.---}Let us consider the following asymmetric two-qubit mixed state~\cite{chen13}
%\begin{widetext}
\begin{equation}\label{avnstate}
\rho_{2}=V|\Psi(\alpha)\rangle\langle\Psi(\alpha)|+(1-V)|\Phi(\alpha)\rangle\langle\Phi(\alpha)|,
\end{equation}
where $|\Psi(\alpha)\rangle= \cos\alpha |HH\rangle + \sin\alpha |VV\rangle$ and $|\Phi(\alpha)\rangle=\sin\alpha|HV\rangle+\cos\alpha|VH\rangle$.
Obviously, $\rho_{2}$ is entangled for the region of
$\alpha\in(0,\pi/2),\;V\in[0,1/2)\cup(1/2,1]$.

For the state Eq. (\ref{avnstate}), we come to compare the performance between the usual 3-setting LSI (Blue line) and the 3-setting GLSI (Red line) in Fig. \ref{AVN-1}. Because the state $\rho_2$ is unchanged under the following operations: $V \rightleftharpoons (1-V)$ and flipping Alice's states (i.e., $|H\rangle \rightleftharpoons|V\rangle$), thus the performance in the region $V\in(1/2,1]$ will be the same as that in the region $V\in[0,1/2)$.
Without loss of generality, we choose the region of  $\alpha\in(0,\pi/2),\;V\in[0,1/2)$, for the usual 3-setting LSI the upper bound $V_{\mathrm{Max}}=\frac{1-\sqrt{3}+2\sin{2\alpha}}{2(1+\sin{2\alpha})}$. It is obvious to conclude that there are no states violate the usual 3-setting LSI with the range of $\alpha\in[0,\frac{\arcsin{\frac{\sqrt{3}-1}{2}}}{2}\approx 0.1873]$. However, the 3-setting GLSI can detect some more steerable states for a wider region of $\alpha$ and $V$, which can be calculated numerically. See Fig.~\ref{AVN-1} and Fig.~\ref{AVN-2}, for example, when $\alpha=\frac{\arcsin{\frac{\sqrt{3}-1}{2}}}{2} (\approx\frac{\pi}{17})$, there are no states violate the usual LSI, however, the GLSI still can detect quantum states in the region $V\in [0, V_{\mathrm{Max}}\approx 0.0889]$. On the range of $\alpha\in[\frac{\arcsin{\frac{\sqrt{3}-1}{2}}}{2},\alpha_{B}]$, $\alpha_{B}\approx0.4597\approx\frac{5\pi}{34}$, there exist some states violate the usual 3-setting LSI, but the upper bound $V_{\mathrm{Max}}$ is lower than that of the 3-setting GLSI. On the range of $\alpha\in[\alpha_{B},\frac{\pi}{4}]$, both inequalities are almost equivalent in the task of steering test. When $\alpha=\frac{\pi}{4}$, both of them achieve $V_{\mathrm{Max}}=\frac{3-\sqrt{3}}{4}$.

\section{Appendix C. Experimental details.}
A 404nm laser is sent into a nonlinear crystal BBO to generate maximally entangled state of the form $|\phi\rangle=\frac{1}{\sqrt{2}}(|H\rangle_{A} |V\rangle_{B}-|V\rangle_{A} |H\rangle_{B})$ with average fidelity over $99\%$. By setting HWP1 at $0^{o}$, the photon of Bob passes through BD1, which splits photon into two paths, upper~($u$) path and lower~($l$) path, according to its polarization, either vertical (V) and horizontal (H). If HWP2 is rotated by an angle $\frac{\beta}{2}$ and HWP3 is fixed at $45^{o}$, the two-photon entangled state becomes

\begin{equation}
|\Psi\rangle=\frac{\sin\beta}{\sqrt{2}}|H\rangle_{A} |H\rangle_{Bu}-\frac{\cos\beta}{\sqrt{2}}|H\rangle_{A} |V\rangle_{Bu}+\frac{1}{\sqrt{2}}|V\rangle_{A} |V\rangle_{Bl},
\end{equation}
where $u$ and $l$ donate upper path and lower path of Bob's photon, respectively.
After that Bob's photon passes through BD2, the V-polarized element of the entangled state is lossed in the upper path. One can verify that the matrix of the process of the asymmetric loss interferometer is 
\begin{equation}
\begin{pmatrix}
     sin\beta & 0 \\
     0 & 1 
\end{pmatrix},
\end{equation}

Therefore, the final state (unnormalized state) is given by

\begin{figure*}
[ht]
\begin{center}
\includegraphics[width=0.9\textwidth]{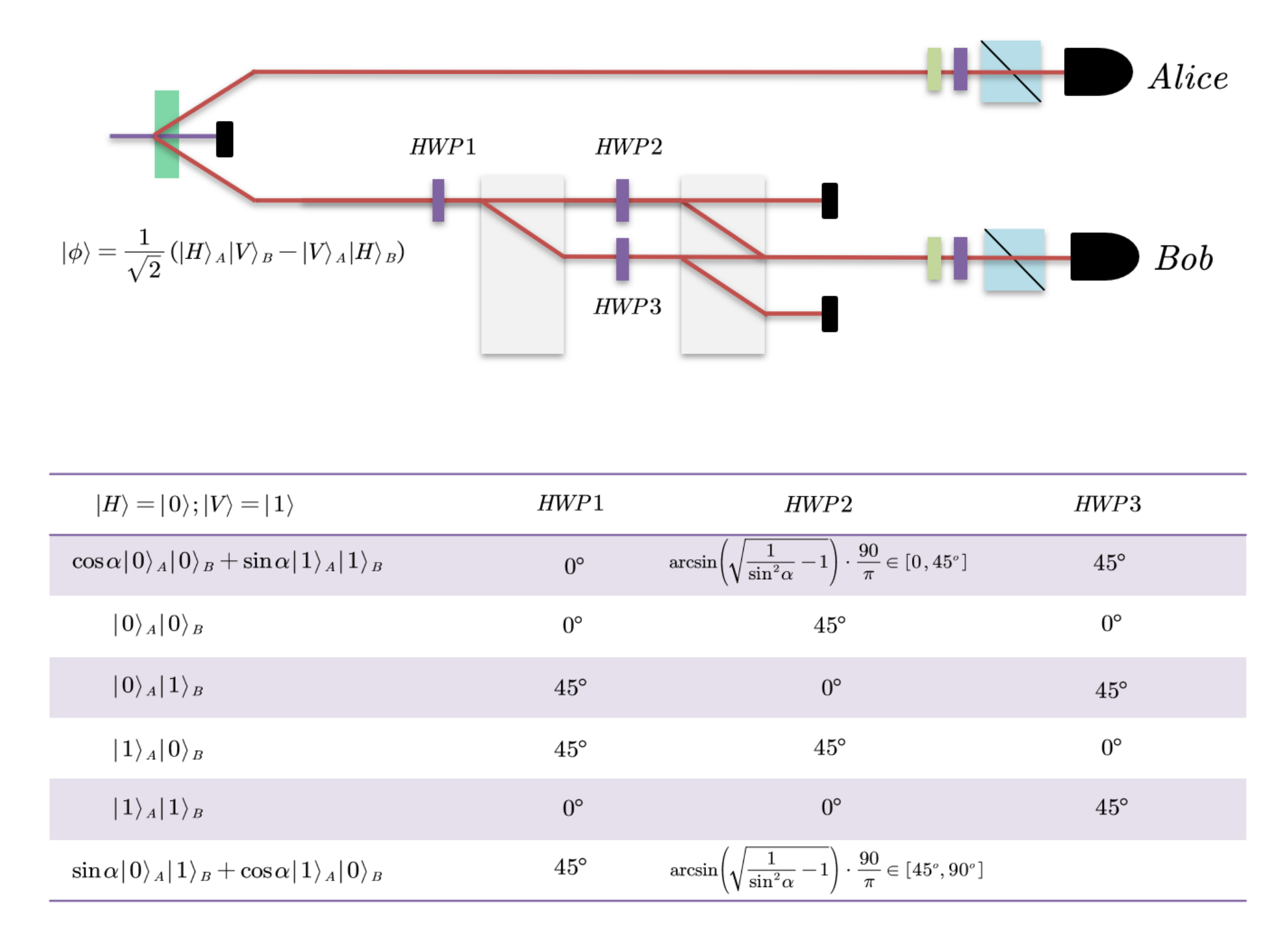}
\vspace{-10 pt}\caption{\label{methods} The experimental setup and the specific angles for state preparation.}%
\end{center}
\end{figure*}

\begin{equation}
|\Psi\rangle=\frac{\sin\beta}{\sqrt{2}}|H\rangle_{A} |H\rangle_{B}+\frac{1}{\sqrt{2}}|V\rangle_{A} |V\rangle_{B}.
\end{equation}
After normalization, the two-qubit entangled state becomes
\begin{equation}
|\Psi\rangle=\frac{\sin\beta}{\sqrt{(\sin\beta)^{2}+1}}|H\rangle_{A} |H\rangle_{B}+\frac{1}{\sqrt{(\sin\beta)^{2}+1}}|V\rangle_{A} |V\rangle_{B}.
\end{equation}
Compared with the form in the main text, by rotating HWP2 by $\beta=\arcsin(\sqrt{\frac{1}{(\sin\alpha)^2}-1})\cdot\frac{90}{\pi}\in[0,45^{o}]$,  we can generate
\begin{equation}
|\Psi\rangle=\cos\alpha|H\rangle_{A} |H\rangle_{B}+\sin\alpha|V\rangle_{A} |V\rangle_{B}.
\end{equation}

In our experiments, the verification of the mixed state is achieved by probabilistically mixing the corresponding pure states. Specifically, we measured the corresponding observables in different pure states, and post-processed the data (change the probability of these pure states and mixing them together) to obtain experimental data of different mixed states.
Now we show how to construct two types of mixed states, the generalized Werner state and a asymmetric mixed state~\cite{chen13},
\begin{eqnarray}
\rho_{1}=V|\Psi\rangle\langle\Psi|+(1-V)\frac{\openone\otimes\openone}{4},\\
\rho_{2}=V|\Psi\rangle\langle\Psi|+(1-V)|\Phi\rangle\langle\Phi|,
\end{eqnarray}
where $|\Phi\rangle=\sin\alpha|H\rangle_{A} |V\rangle_{B}+\cos\alpha|V\rangle_{A} |H\rangle_{B}$, and $V\in[0,1]$ is the visibility (probability for $|\Psi\rangle\langle\Psi|$). 
The preparation of the maximally mixed state $\openone\otimes\openone$ is simulated by mixing the four states $|HH\rangle$, $|HV\rangle$, $|VH\rangle$, and $|VV\rangle$ with equal probability. The generalized Werner state is simulated by mixing $\openone\otimes\openone$ and $|\Psi\rangle\langle\Psi|$ with probability $V$ and $1-V$ respectively. The asymmetric mixed state is simulated by mixing $|\Psi\rangle$ and $|\Phi\rangle$ with probability $V$ and $1-V$ respectively. 
The specific rotation angles of HWP in the beam displacer interferometer for preparation theses quantum states are shown in the table of Fig.~\ref{methods}.

\section{Funding}
This work was supported by National Key R$\&$D Program of China ( 2017YFA0305200,  2016YFA0301300), The Key R$\&$D Program of Guangdong Province ( 2018B030329001, 2018B030325001), The National Natural Science Foundation of China (61974168,  12075245, 11875167, 12075001), and Xiaoxiang Scholars Programme of Hunan Normal university.

\section{Disclosures}
The authors declare that there are no competing interests. 
T. Feng, C. R and Q. F contributed equally to this work.

%\bibitem{DIQKD}
%Y. Zhao, H.Ku, S. Chen, H. Chen, F. Nori, G. Xiang, C. Li, G. Guo and Y. Chen, 
%``Experimental demonstration of measurement-device-independent measure of quantum steering,''
%npj Quantum Inf. \textbf{6}, 77 (2020)

%\bibitem{Bian}
%Z. Bian, A. S. Majumdar, C. Jebarathinam, K. Wang, L. Xiao, X. Zhan, Y. Zhang, and P. Xue,
%``Experimental demonstration of one-sided device-independent self-testing of any pure two-qubit entangled state,''
%Phys. Rev. A \textbf{101} 020301 (2020)


\begin{thebibliography}{99}
\bibitem{X}
For example, the maximal violation value for the usual 3-setting steering inequality (\ref{si-3-2}) is
 $1+2\sin{2\alpha}$. Hence, only when $\alpha>\frac{\arcsin{\frac{\sqrt{3}-1}{2}}}{2}\approx 0.1873$, the usual linear steering inequality can be violated. However, the pure state violates the GLSI (\ref{si-3-1 m}) for the whole region $\alpha\in (0, \pi/2)$, thus the GLSI is stronger than the usual LSI in detecting the EPR steerability of pure entangled states.
\end{thebibliography}
\end{document}